\documentclass[journal]{IEEEtran}
\usepackage{array}
\usepackage{amsmath,amsfonts}
\usepackage{algorithmic}
\usepackage{algorithm}
\usepackage{array}
\usepackage[caption=false,font=normalsize,labelfont=sf,textfont=sf]{subfig}
\usepackage{textcomp}
\usepackage{stfloats}
\usepackage{url}
\usepackage{verbatim}
\usepackage{graphicx}
\usepackage{cite}
\usepackage{amssymb}
\usepackage{color}
\usepackage{booktabs}
\usepackage{makecell}
\usepackage{pifont}
\usepackage{rotating}
\usepackage{multirow}
\usepackage{siunitx} 
\usepackage{booktabs}
\usepackage{dsfont}
\begin{document}

\title{Beam Prediction Based on Multimodal Large Language Models\\
% \thanks{Identify applicable funding agency here. If none, delete this.}
}
\author{Tianhao Mao, \IEEEmembership{Student Member,~IEEE},
        Le Liang, \IEEEmembership{Member,~IEEE},  
        Jie Yang, \IEEEmembership{Member,~IEEE}, \\
        Xiao Li, \IEEEmembership{Member,~IEEE},
        Shi Jin, \IEEEmembership{Fellow,~IEEE},
        and Geoffrey Ye Li, \IEEEmembership{Fellow,~IEEE}
% \thanks{The work was supported in part by the National Key Research and Development Program of China under Grant 2024YFE0200103, in part by the National Natural Science Foundation of China (NSFC) under Grant 62301156. The work was also supported by the Fundamental Research Funds for the Central Universities 2242022K60004 and the Key Technologies R\&D Program of Jiangsu (Prospective and Key Technologies for Industry) under Grants BE2023022 and BE2023022-1. (\emph{Corresponding author: Le Liang; Jie Yang}).}
\thanks{T. Mao, L. Liang, X. Li, and S. Jin are with the School of Information Science and Engineering, Southeast University, Nanjing 210096, China (e-mail: tianhao@seu.edu.cn; lliang@seu.edu.cn; li\_xiao@seu.edu.cn; jinshi@seu.edu.cn). L. Liang is also with the Purple Mountain Laboratories, Nanjing 211111, China.}% <-this % stops a space
% \thanks{Y. Chen is with Chien-Shiung Wu College, Southeast University, Nanjing 210096, China (e-mail: 213241319@seu.edu.cn).}
\thanks{J. Yang is with the Key Laboratory of Measurement and Control of Complex Systems of Engineering, Ministry of Education, Southeast University, Nanjing 210096, China (e-mail: yangjie@seu.edu.cn).}
% \thanks{H. Ye is with the Department of Electrical and Computer Engineering, University of California, Santa Cruz, CA 95064, USA (e-mail:
 % hye30@ucsc.edu).}
\thanks{G. Y. Li is with the ITP Lab, the Department of Electrical and Electronic Engineering, Imperial College London, SW7 2BX London, U.K.
 (e-mail: geoffrey.li@imperial.ac.uk).}
 \thanks{This paper was accepted in part by the 2026 IEEE International Conference on Communications \cite{icc2026}.}
 }

\maketitle

\begin{abstract}
Accurate beam prediction is a key enabler for next-generation wireless communication systems. In this paper, we propose a beam prediction framework based on multimodal large language models (LLMs). It effectively utilizes contextual information provided by sensory data including RGB camera images and LiDAR point clouds. To effectively fuse heterogeneous modalities, we design specialized modality encoders together with a beam-guided attention masking mechanism and a high-frequency temporal alignment strategy, enabling robust cross-modal feature integration under dynamic environments. Furthermore, we construct a large-scale multimodal dataset for communication, named Multimodal-Wireless, which covers diverse weather and traffic conditions with high-fidelity ray-tracing labels. Extensive simulation results demonstrate that the proposed approach significantly reduces the reliance on oracle angle of departure knowledge and consistently outperforms state-of-the-art multimodal LLM-based beam prediction methods in terms of beam accuracy and communication performance, improving the average Top-1 accuracy to 80.8\% and the average normalized gain to 89.1\%.
% Leveraging contextual sensing information to optimize spatial transmission is a cornerstone of next-generation communication systems. In this study, we propose a novel beam prediction framework that harnesses the power of multimodal large language models (LLMs). Unlike state-of-the-art approaches that heavily rely on historical angle of departure (AoD) information—which is typically unavailable at the base station—our framework operates exclusively on readily available on-site sensory data. We utilize specialized modality encoders for RGB camera and light detection and ranging (LiDAR), integrating a proposed beam-guided attention masking mechanism and a high-frequency temporal alignment strategy to fuse features effectively. In addition, we build a large-scale dataset named Multimodal-Wireless, which covers diverse weather and traffic conditions with high-fidelity ray-tracing data. Simulation results demonstrate that leveraging multimodal sensing information successfully replaces the reliance on perfect AoD knowledge and outperforms the state-of-the-art LLM-based algorithm.}

\end{abstract}


\section{Introduction}

\IEEEPARstart{T}{he} performance of future communication systems, particularly those employing massive multiple-input multiple-output (MIMO) at high frequencies, hinges on precise beam alignment. However, maintaining this alignment under user mobility presents a fundamental dilemma. Conventional methods typically rely on exhaustive beam sweeping or reactive recovery protocols, which incur prohibitive latency and consume substantial signaling overhead. Such heavy training burdens severely degrade spectral efficiency and often lead to service interruptions when the channel changes faster than the feedback loop. To break these bottlenecks, a paradigm shift from reactive to proactive management is essential, necessitating a dynamic understanding of the operational context derived from environmental sensing \cite{liufov}. By leveraging such context to anticipate optimal beams before link degradation occurs, proactive beam prediction becomes a key enabler for reducing training overhead and ensuring seamless connectivity in next-generation networks.

To achieve such proactivity, context-aware communication strategies have been proposed to mitigate the overhead inherent to millimeter-wave (mmWave) systems \cite{TMC, ckm, bpllm}. While some strategies rely on user feedback or assume perfect knowledge of historical angle of departure (AoD) \cite{bpllm}, such information is often unavailable or delayed at the base station (BS). In contrast, local sensory data derived from on-site RGB cameras and light detection and ranging (LiDAR) is readily accessible and provides rich geometric and visual cues about the propagation environment. Hence, {in this paper, we leverage} exclusively BS-side sensing ata as a practical and deployable solution for robust context-aware prediction.

{There have been several studies on deep learning-based physical layer techniques leveraging such multimodal data} \cite{wcnc, iot, shibingpu, cuiyuanhao, qinzhijin, heath}. {From \cite{wcnc, shibingpu},} using visual, positional, and LiDAR data {can} accelerate beam selection or predict optimal beams. Attention mechanisms {have been further introduced} to fuse these heterogeneous features \cite{cuiyuanhao}. However, due to their limited parameter scale and architectural constraints, these traditional learning-based models, like long short-term memory (LSTM), often struggle to capture complex spatial-temporal dependencies. Consequently, existing small-scale learning-based predictors often fail to generalize effectively under dynamic environmental shifts and unseen scenarios \cite{LLMoverview}.

To address generalization issues, recent work \cite{bpllm} has adopted large language models (LLMs) for beam prediction, demonstrating significantly improved robustness compared to conventional methods. However, this approach relies on historical channel angles rather than raw sensory data. Furthermore, simply applying existing multimodal LLMs—from foundational encoders like CLIP \cite{clip} and Imagebind \cite{imagebind} to systems like BLIP-2 \cite{blip} and Imagebind-LLM \cite{imagebind_llm}—is insufficient. Current multimodal paradigms are designed to align semantically similar cross-modal data, whereas wireless applications require exploiting the heterogeneity where LiDAR, cameras, and channels carry {different and} complementary information. Therefore, we need a specialized multimodal LLM framework that performs complementary fusion, which inherently necessitates matched high-frequency, weather-diverse, and BS-side aligned training data.

Meanwhile, the development of such data-driven models depends heavily on specialized datasets. While {there exist} sensing-centric datasets, like OPV2V \cite{opencood} and DAIR-V2X \cite{dair-v2x}, and communication-focused ones, like DeepMIMO \cite{deepmimo}, bridging {them} remains difficult. Efforts, such as e-Flash \cite{e-flash}, ViWi \cite{viwi} and DeepSense 6G \cite{DeepSense}, have begun to integrate these domains, yet persistent limitations remain, including the omission of adverse weather conditions, a reliance on low-frequency received power, or a lack of precise ray-tracing alignment. These gaps hinder the training of models that can withstand realistic environmental impairments like rain or fog.
% To this end, we introduce Multimodal-Wireless to bridge these gaps and facilitate robust model development.

% In this paper, we propose a novel multimodal LLM-based beam prediction framework that operates without relying on historical AoD information. We design specialized modality encoders with a beam-guided attention masking (BGAM) mechanism to resolve spatial ambiguity and introduce a high-frequency temporal alignment strategy to fuse heterogeneous sensory data. Furthermore, we construct the large-scale Multimodal-Wireless dataset, covering diverse weather and traffic conditions with high-fidelity ray-tracing labels. Extensive simulation results demonstrate that the proposed approach significantly reduces the reliance on oracle AoD knowledge and consistently outperforms state-of-the-art multimodal LLM-based beam prediction methods in terms of beam accuracy and communication performance.

In this paper, we address these key challenges by proposing a novel multimodal LLM framework that operates exclusively on readily available, BS-side sensing data. Our architecture utilizes modality-specific encoders and a unique fusion interface that preserves high-frequency channel information by upsampling sensory features. Trained and tested on the uniquely developed Multimodal-Wireless dataset, our model significantly outperforms the state-of-the-art approach. Our main contributions are as follows:
\begin{itemize}
\item We propose a multimodal LLM framework for robust beam prediction, integrating a beam-guided attention mechanism (BGAM) and high-frequency temporal alignment. BGAM uses historical beam indices to guide LiDAR extraction, resolving spatial ambiguity, while the alignment upsamples sensory features to match 100 Hz channel data.

\item We introduce Multimodal-Wireless, an extensible open-source dataset for joint sensing and communication. It provides 100 Hz channel state information and five sensor modalities across diverse weather conditions, enabling robustness evaluations while allowing researchers to easily generate, align, and contribute custom data.

\item Integrating multimodal data (particularly LiDAR) with beam indices significantly improves long-term prediction. By eliminating impractical reliance on historical AoD, our model achieves higher normalized gain than a state-of-the-art LLM baseline, providing a viable and robust solution for real-world deployment.
\end{itemize}

The rest of the paper is organized as follows. In Section~\ref{sec:problem}, we first formulate the long-term beam prediction problem. We then clarify the network structure in Section~\ref{sec:network}, including the modality encoder, modality interface, and output projection. The Multimodal-Wireless dataset for training and testing the model is introduced in Section~\ref{sec:dataset}. In Section~\ref{sec:simulation}, we present comprehensive simulation results related to multi-modality, weather conditions, and robustness. Finally, we conclude the paper in Section~\ref{sec:conclusion}.

The notations used {in this paper} are defined as follows. $\mathbf{a}$ and $\mathbf{A}$ denote a column vector and a matrix, respectively, and $\mathbf{A}[i,j]$ refers to the entry at the $i$-th row and $j$-th column of $\mathbf{A}$. The transpose and conjugate transpose of $\mathbf{A}$ is denoted by $\mathbf{A}^{\mathrm{T}}$ and $\mathbf{A}^{\mathrm{H}}$, respectively. The floor of a real number $x$, denoted by $\lfloor x \rfloor$, is the greatest integer smaller than or equal to $x$. $|y|$ represents the {magnitude} of a complex number $y$.

\section{Problem Formulation} \label{sec:problem}
\begin{figure}[t]
	\centerline{\includegraphics[width=3.4in]{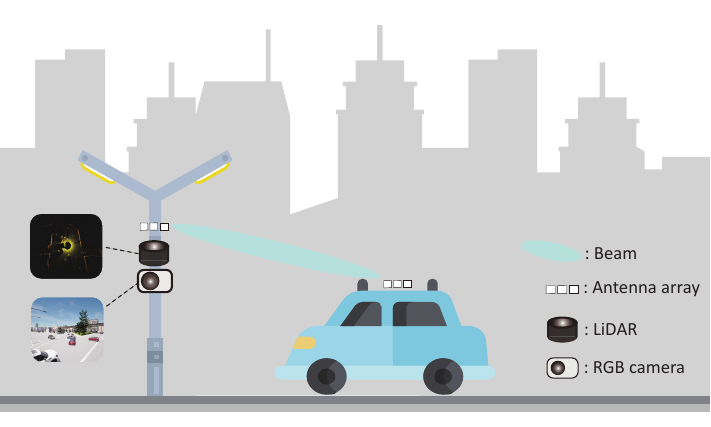}}
	\caption{Illustration of beam prediction in MIMO communications with multimodal data.}
	\label{fig:systemmodel}
\end{figure}

We consider a dynamic downlink system where a roadside unit (RSU) acts as the BS transmitter, and {the user receiver is inside a vehicle}, as illustrated in Fig.~\ref{fig:systemmodel}. 
The RSU is equipped with an $N_{\mathrm{t}}$-antenna uniform linear array (ULA), and the {user} with an $N_{\mathrm{r}}$-antenna ULA. The inter-element spacing, $d$, for {adjacent} arrays is set as $\lambda/2$, where $\lambda$ is the carrier wavelength. In addition, the RSU collects local sensory data, among which it utilizes LiDAR and RGB camera data for physical layer design.
% The physical channel propagation is simulated using the Sionna 2D ray-tracing engine \cite{sionna}.
The communication setup yields a geometric multi-path channel whose frequency-domain representation for the $k$-th subcarrier at time step $n$ is expressed as \cite{yangjie}
\begin{equation} \label{eq:channel_model}
    \!\!\!\!\mathbf{H}_{n,k}\!=\!\sqrt{\frac{N_\mathrm{t}N_\mathrm{r}}{L_n}}\sum_{l=1}^{L_n}\alpha_{n,l}\mathbf{a}_{\mathrm{r}}(\theta_{n,l})\mathbf{a}_{\mathrm{t}}^{\mathrm{H}}(\phi_{n,l})e^{-j2\pi f_k\tau_{n,l}},
\end{equation}
{where} $L_n$ represents the number of multi-paths at time step $n$, {and e}ach path is characterized by a complex gain $\alpha_{n,l}$, delay $\tau_{n,l}$, angles of arrival (AoA) $\theta_{n,l}$, and AoD $\phi_{n,l}$. For an orthogonal frequency division multiplexing (OFDM) system with $K$ subcarriers and $\Delta f$ subcarrier spacing, $f_k$ is the baseband frequency of the $k$-th subcarrier and satisfies $f_k=(k-\frac{K+1}{2})\Delta f$. The steering vectors are defined as
\begin{equation} \label{eq:steeringvector}
\begin{aligned}
    \mathbf{a}_{\mathrm{t}}(\theta) &= \left[1,\, e^{j 2\pi d \sin\theta / \lambda},\, \cdots,\, e^{j 2\pi (N_{\mathrm{t}}-1) d \sin\theta / \lambda} \right]^{\mathrm{T}}, 
        \\
        \mathbf{a}_{\mathrm{r}}(\phi) &= \left[1,\, e^{j 2\pi d \sin\phi / \lambda},\, \cdots,\, e^{j 2\pi (N_{\mathrm{r}}-1) d \sin\phi / \lambda} \right]^{\mathrm{T}}.
\end{aligned}
\end{equation}
% This model aligns with the general Sionna channel formulation in \eqref{eq:Sionna_channel}, where the per-path matrix $\mathbf{A}_m$ encapsulates the path gain and spatial signatures:
% \begin{equation}
%     \mathbf{A}_m=\sqrt{\frac{N_\mathrm{t}N_\mathrm{r}}{L_n}}\alpha_m\mathbf{a}_{\mathrm{r}}\left(\theta_{\mathrm{r},m}\right)\mathbf{a}_{\mathrm{t}}^{\mathrm{H}}\left(\theta_{\mathrm{t},m}\right).
% \end{equation}
The system sends a single data stream with a precoder $\mathbf{f}\in\mathbb{C}^{N_{\mathrm{t}}\times1}$ at the RSU and a combiner $\mathbf{w}\in\mathbb{C}^{N_{\mathrm{r}}\times1}$ at the {user}. Thus, the received signal can be written as
\begin{equation}
    y_{n,k} = \mathbf{w}^\mathrm{H}\mathbf{H}_{n,k} \mathbf{f}x_{n,k}+s_{n,k},
\end{equation}
where $x_{n,k}$ and $y_{n,k}$ are the signal transmitted and received at the $n$-th time step and the $k$-th subcarrier, and $s_{n,k}$ is the communication noise, an independent and identically distributed
(i.i.d.) circularly symmetric complex Gaussian random variable with zero mean and a variance of $\sigma^2$.

We consider an analog beamforming architecture with a single radio frequency chain, with both $\mathbf{f}$ and $\mathbf{w}$ selected from a predefined discrete {F}ourier transform (DFT) codebook.
For a codebook of size $Q$, the $q$-th candidate beamformer is expressed as
\begin{equation} \label{eq:codebook}
    \mathbf{f}(q)=\frac{1}{\sqrt{N_\mathrm{t}}}
    \begin{bmatrix}
        1, & e^{j2\pi q/Q}, & \cdots, & e^{j2\pi(N_{\mathrm{t}}-1)q/Q}
    \end{bmatrix}^{\mathrm{T}}.
\end{equation}

The primary task addressed in this work is to predict the optimal transmit beam indices for the next $W$ time steps, $\{\hat{{q}}_m\}_{m=1}^W$, 
% $\hat{\mathbf{q}}_W=[\hat{q}_1,\dots,\hat{q}_W]^{\mathrm{T}}$, with $\hat{q}_n$ being the predicted beam index at time step $n$, 
% based on multimodal data from the previous $P$ time steps. This data includes historical beam indices $\{q^*_n\}_{n=1}^P$, LiDAR point clouds, and camera images. 
based on historical information from the past $P$ time steps, including optimal transmit beam indices, LiDAR point clouds, and camera images.
Mathematically, we aim to find the series of beam indices $\{\hat{q}_1,\dots,\hat{q}_m,\dots\}$ that achieve the maximum sum of normalized beamforming gain for future $W$ time steps, which is expressed by
\begin{equation}
    \max_{\{\hat{{q}}_m\}_{m=1}^W}\frac{1}{W}\sum_{m=1}^{W} G(\hat{q}_m),
\end{equation}
where $G(\hat{q}_m)$ is the normalized gain of the future time step $m$, defined by
\begin{equation} \label{eq:normalizedgain}
    G(\hat{q}_m)=\frac{|\mathbf{w}(p_m^*)^{\mathrm{H}}\mathbf{H}_m \mathbf{f}(\hat{q}_m)|^2}{|\mathbf{w}(p_m^*)^{\mathrm{H}}\mathbf{H}_m \mathbf{f}(q_m^*)|^2},
\end{equation}
with $\mathbf{H}_m = \frac{1}{K}\sum_{k=1}^{K}\mathbf{H}_{m,k}$ representing the channel matrix averaged over all $K$ subcarriers. Notably, due to the hardware constraints of the analog beamforming architecture, a common beamformer and combiner are applied across all subcarriers. {Nevertheless, our beam prediction method can be directly extended to the case where per-subcarrier processing is performed at the digital baseband on the equivalent channel.} $p_m^*$ and $q_m^*$ denote the optimal precoder and combiner beam indices at future time step $m$,  respectively, determined via an exhaustive search over the codebook to maximize the received signal power:
\begin{equation} \label{eq:optimal_beam}
    (p_m^*, q_m^*) = \arg\max_{p_m, q_m \in \{0, \dots, Q-1\}} \left|\mathbf{w}(p_m)^{\mathrm{H}} \mathbf{H}_m \mathbf{f}(q_m)\right|^2.
\end{equation}

\section{Multimodal LLM-based beam prediction} \label{sec:network}
\begin{figure*}[t]
	\centerline{\includegraphics[width=6.6in]{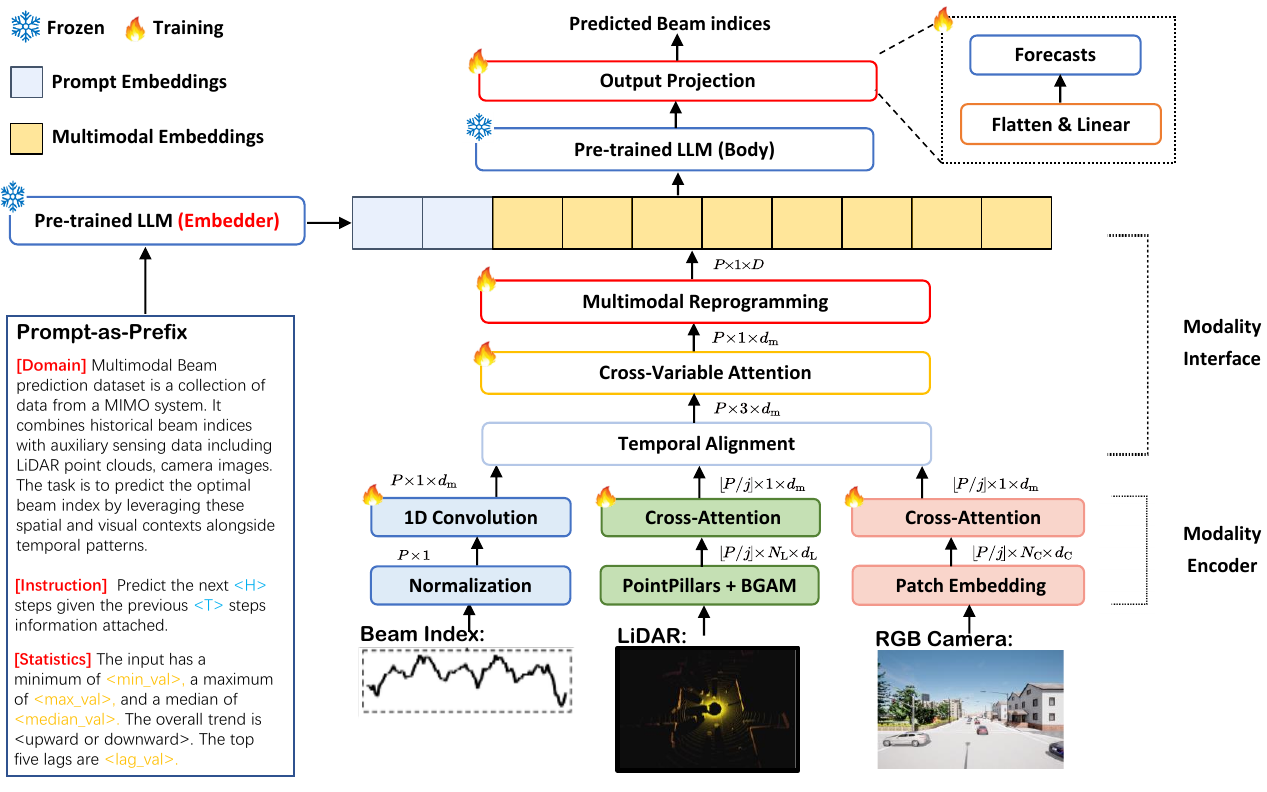}}
	\caption{The structure of the multimodal LLM-based beam prediction network.}
	\label{fig:network}
\end{figure*}

% The Multimodal-Wireless dataset offers a comprehensive collection of multimodal data samples, encompassing both traditional sensor readings and channel matrices derived from a ray-tracing model. A key feature of this dataset is that all modalities inherently capture time-varying geometric information. Utilizing this rich dataset, primarily from the RSU side, we train an MLLM to tackle the task of long-term beam prediction for downlink communication.
In this section, we introduce our multimodal LLM-based method, comprising two key components: a modality encoder for feature extraction of each data type, and a modality interface for temporal alignment and subsequent fusion of these multimodal features. We note that by reprogramming an embedding-visible LLM, the optimal transmit beam prediction can be performed without fine-tuning the backbone model \cite{bpllm}. The structure of our method is presented in Fig.~\ref{fig:network}.

\subsection{Modality Encoder} \label{subsec:modalityencoder}
Unlike traditional encoders that align the semantic space of different modalities~\cite{clip}, our approach preserves the intrinsic heterogeneity of each modality. This is designed to exploit complementary information, allowing the strengths of one modality to compensate for the weaknesses of another. To achieve this, we design modality-specific encoders for heterogeneous modality data to extract the diverse features for the beam prediction task, ensuring tailored processing for each modality before its features are interpreted by the LLM.

\subsubsection{Beam Index} 
% The optimal beam indices for training are calculated with \eqref{eq:optimal_beam} on the basis of channels in the Multimodal-Wireless dataset. 
We use past optimal beam indices $\mathbf{q}^*_P\in \mathbb{R}{}^{P \times 1}
$ from the BS side as historical observations.
Given that different codebook sizes lead to distinct ranges of beam indices, we adopt the pre-processor in \cite{bpllm} to normalize the past optimal beam index sequence $\mathbf{q}^*_P$ by 
        \begin{equation}
            \mathbf{q}^*=\frac{\mathbf{q}^*_P}{Q},
        \end{equation}
    in order to enhance the generalization of our model.
    
    For feature embedding, we adopt an embedding layer to map $\mathbf{q}^*\in\mathbb{R}^{P\times 1}$ into $\mathbf{U}_{\mathrm{B}}\in\mathbb{R}^{P\times d_\mathrm{m}}$, where $d_{\mathrm{m}}$ is the unified dimension within the modality encoding stage, preparing the features for the subsequent modality interface, introduced in Section \ref{sec:interface}. Specifically, the layer uses a one-dimensional convolution to project input features into a target embedding space, and processes the input sequence $\mathbf{q}^{*}$ with $d_\mathrm{m}$ convolution kernels of size 3, which captures local contextual information from neighboring time steps when creating the embedding for each time step.
    %The use of circular padding suggests it is particularly well-suited for handling time-series data.

\subsubsection{LiDAR} The raw LiDAR point cloud, comprising sparse and unordered points within the three-dimensional (3D) perception space, is processed by a Pointpillar network \cite{pointpillar}, a self-designed beam-guided attention masking (BGAM) and a cross-attention layer in sequence.

\begin{figure}[t]
	\centerline{\includegraphics[width=3.4in]{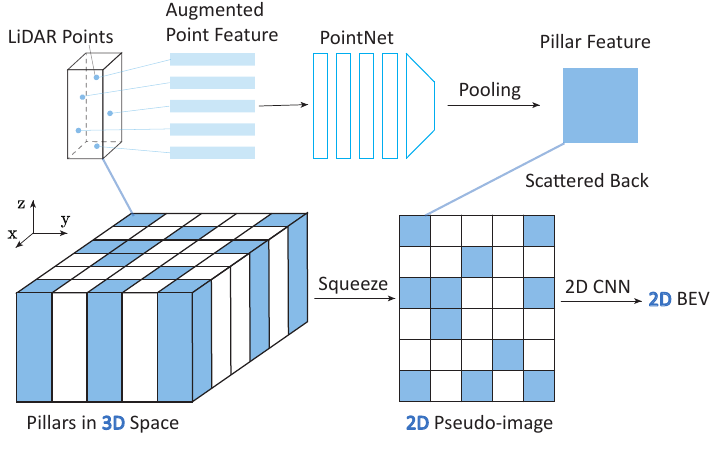}}
	\caption{The 3D to 2D BEV process of Pointpillar \cite{pointpillar}.}
	\label{fig:pointpillar}
\end{figure}

\textbf{Pointpillar:} The point cloud is first processed by a frozen pre-processor that adopts the first two stages of the PointPillars architecture, i.e., a feature encoder and a 2D convolutional backbone, transforming the point cloud into a sparse bird’s eye view (BEV).
% Specifically, the point cloud is first discretized into grids in the $x-y$ plane, forming pillars. 
The process is illustrated in Fig. \ref{fig:pointpillar}. Specifically, the 3D perception space of LiDAR is first discretized into vertical pillars by gridding the $x-y$ plane.
% For each non-empty pillar, the internal points are augmented to a 9-dimensional feature, incorporating offsets from the pillar's center and the points' centroid. To handle the inherent data sparsity, a fixed-size dense tensor is generated by sampling a maximum number of non-empty pillars and points per pillar. This tensor is processed by PointNet~\cite{pointnet} to learn point-level features. A subsequent max-pooling operation across the point dimension aggregates these into a pillar feature tensor. All pillar features are then scattered back to their original 2D grid locations to create a pseudo-image. This pseudo-image is processed by a 2D convolutional neural network, which includes a top-down sub-network for producing features at decreasing spatial resolutions and an upsampling sub-network that concatenates them into a multi-scale feature representation. 
Points within each non-empty pillar are augmented into 9-dimensional features by incorporating offsets from the pillar's center and the points' centroid. To mitigate inherent data sparsity, sampling a bounded number of pillars and points yields a fixed-size dense tensor. PointNet~\cite{pointnet} then extracts point-level features, which are subsequently aggregated into a unified pillar representation via point-wise max-pooling. By scattering these features back to their original 2D grid locations, a pseudo-image is constructed. Finally, a 2D convolutional neural network processes this pseudo-image, employing a top-down pathway to extract decreasing-resolution features and an upsampling module to concatenate them into a multi-scale representation.

We note here that the weights for these two stages are initialized from a public implementation~\cite{opencood} and remain frozen during training. The output of this pre-processing is the BEV feature map $\mathbf{X}_{\mathrm{L}}\in\mathbb{R}^{C_{\mathrm{L}}\times H_\mathrm{L} \times W_\mathrm{L}}$.

\textbf{Beam-Guided Attention Masking:}
A key challenge in multimodal beam prediction is bridging the semantic gap between the abstract communication modality, i.e., discrete beam indices, and the rich spatial information from sensory data. To address this, we introduce a BGAM module that injects inductive bias by enforcing the physical constraints of the communication channel directly onto the BEV feature. This acts as a hard attention mechanism, guiding the model to focus solely on the spatial region relevant to a specific beam.

The BGAM operates as follows. We first map each historical beam index $q \in \{0, \dots, Q-1\}$ to its corresponding continuous physical AoD, $\phi_q$. 
%% MODIFIED HERE: The following paragraph is revised to align with the paper's equations.
This mapping is derived by aligning the phase progression of the $n$-th antenna element in the $q$-th DFT beamformer, as defined in \eqref{eq:codebook}, with the phase of the physical steering vector for a ULA, as defined in \eqref{eq:steeringvector}. Assuming a standard inter-element spacing of half a wavelength , i.e., $d=\lambda/2$, this alignment yields a direct relationship between the beam index and its pointing angle. To ensure the generated beams cover the full angular spectrum from $-90^\circ$ to $+90^\circ$, the codebook is designed such that the sine of the angle is sampled uniformly in the range $[-1, 1]$. This results in the following mapping:
\begin{equation}
    \sin(\phi_q) = \frac{2(q+0.5)}{Q} - 1.
    \label{eq:angle_mapping}
\end{equation}
From this, the center angle for each beam is given by 
\begin{equation}
    \phi_q = \arcsin\left(\frac{2(q+0.5)}{Q} - 1\right).
\end{equation}

To generate the spatial mask, we define an angular range $[\phi_q - \delta_q, \phi_q + \delta_q]$, where $\delta_q$ is the corresponding adaptive angular half-width, which satisfies
\begin{equation}
    \sin(\phi_q + \delta_q) - \sin(\phi_q - \delta_q) \propto \frac{2}{Q}.
\end{equation}

Concurrently, we compute a static angle grid, $\boldsymbol{\Theta} \in \mathbb{R}^{H_\mathrm{L} \times W_\mathrm{L}}$, where each element $\boldsymbol{\Theta}[i,j]$ stores the pre-calculated angle of the center of the BEV grid cell $(i,j)$ relative to the RSU. A binary spatial mask $\mathbf{M} \in \{0, 1\}^{H_\mathrm{L} \times W_\mathrm{L}}$ is then generated for the beam index $q$ according to,
\begin{equation}
    \mathbf{M}[i,j] = 
    \begin{cases} 
        1, & \text{if } \boldsymbol{\Theta}[i,j] \in [\phi_q - \delta_q, \phi_q + \delta_q], \\
        0, & \text{otherwise.}
    \end{cases}
\end{equation}
This mask isolates the spatial region covered by the beam.
The high-level BEV feature map $\mathbf{X}_\mathrm{L}$ can be viewed as a grid of feature vectors, where $\mathbf{x}_{i,j} \in \mathbb{R}^{C_\mathrm{L}}$ denotes the feature representation at the spatial location $(i,j)$. The masking operation is applied spatially as,
\begin{equation}
    \hat{\mathbf{x}}_{i,j} = \mathbf{x}_{i,j} \cdot \mathbf{M}[i,j].
    \label{eq:feature_masking}
\end{equation}
Here, the scalar value, $\mathbf{M}[i,j] \in \{0, 1\}$, acts as a gate, either preserving or suppressing the entire feature vector at each grid cell based on the beam's coverage. Subsequently, by spatially stacking the masked feature vectors, $\hat{\mathbf{x}}_{i,j}$, across all grid cells, we reconstruct the refined LiDAR BEV feature map $\hat{\mathbf{X}} \in \mathbb{R}^{C_{\mathrm{L}} \times H_\mathrm{L} \times W_\mathrm{L}}$, which serves as the beam-aware representation for downstream prediction.
By zeroing out features outside the beam's direct line-of-sight (LOS), this BGAM process significantly prunes the feature space, reducing noise and computational complexity for downstream components. More importantly, it forces an explicit alignment between the communication state and the environmental perception, improving the model's interpretability.
% and focusing its learning on the causal relationship between observed obstacles within a beam's path and the resulting channel quality.

\textbf{Cross-Attention Layer:} 
To transform the 2D spatial representation into a format compatible with the attention-based mechanism, the resulting tensor, $\hat{\mathbf{X}}$, is then flattened into $\tilde{\mathbf{X}}_{\mathrm{L}} \in \mathbb{R}^{N_\mathrm{L} \times C_{\mathrm{L}}}$, with $N_\mathrm{L}=H_\mathrm{L}\times W_\mathrm{L}$. We then employ a cross-attention mechanism to summarize these features using a learnable query vector $\mathbf{q}_{\mathrm{L}}\in\mathbb{R}^{1\times C_{\mathrm{L}}}$. The operation is
    \begin{equation}
        \mathbf{u}_{\mathrm{L}}=\mathrm{CrossAttention}(\mathbf{q}_{\mathrm{L}},\tilde{\mathbf{X}}_{\mathrm{L}},\tilde{\mathbf{X}}_{\mathrm{L}}),
        \label{eq:cross_attention}
    \end{equation}
    where $\mathbf{u}_{\mathrm{L}}\in \mathbb{R}^{1\times d_\mathrm{m}}$ is the final embedded feature.

    Our cross-attention is implemented with a standard multi-head mechanism with $N_\mathrm{h}$ heads. First, the query, key, and value are linearly projected using shared weight matrices $\mathbf{W}_{\mathrm{q}}\in\mathbb{R}^{C_{\mathrm{L}}\times d_\mathrm{m}}$, $\mathbf{W}_{\mathrm{k}}\in\mathbb{R}^{C_{\mathrm{L}}\times d_\mathrm{m}}$, and $\mathbf{W}_{\mathrm{v}}\in\mathbb{R}^{C_{\mathrm{L}}\times d_\mathrm{m}}$ as
    \begin{equation}
    \begin{aligned}
        \mathbf{q}' &= \mathbf{q}_{\mathrm{L}}\mathbf{W}_{\mathrm{q}}, \\
        \mathbf{K}' &= \tilde{\mathbf{X}}_{\mathrm{L}}\mathbf{W}_{\mathrm{k}}, \\
        \mathbf{V}' &= \tilde{\mathbf{X}}_{\mathrm{L}}\mathbf{W}_{\mathrm{v}}.
    \end{aligned}
    \end{equation}
    The resulting matrices, $\mathbf{q}'\in\mathbb{R}^{1\times d_\mathrm{m}}$, $\mathbf{K}'\in\mathbb{R}^{N_{\mathrm{L}}\times d_\mathrm{m}}$, and $\mathbf{V}'\in\mathbb{R}^{N_{\mathrm{L}}\times d_\mathrm{m}}$, are then split into $N_\mathrm{h}$ heads. For each head $s\in\{1, \dots, N_\mathrm{h}\}$, the inputs are $\mathbf{Q}_s\in\mathbb{R}^{1\times d}$, $\mathbf{K}_s\in\mathbb{R}^{N_{\mathrm{L}}\times d}$, and $\mathbf{V}_s\in\mathbb{R}^{N_{\mathrm{L}}\times d}$, where $d = d_\mathrm{m}/N_\mathrm{h}$ is the dimension of each head. The attention output for head $s$ can be written as
    \begin{equation}
        \mathbf{u}_s=\mathrm{Softmax}\left(\frac{\mathbf{q}_s\mathbf{K}_s^{\mathrm{T}}}{\sqrt{d}}\right)\mathbf{V}_s.
        \label{eq:single_head_attention}
    \end{equation}
    Finally, the outputs of all heads are concatenated to form
    \begin{equation}
        \mathbf{u}_{\mathrm{L}} = \mathrm{Concat}(\mathbf{u}_1, \dots, \mathbf{u}_{N_\mathrm{h}}),
    \end{equation}
    which is the feature embedding of a single LiDAR point cloud. To align with the dimensions of the beam index embedding, $\mathbf{U}_{\mathrm{B}}$, the encoded features of the undersampled LiDAR point clouds from the past $P$ time steps are replicated and concatenated to construct the historical LiDAR feature embedding $\mathbf{U}_{\mathrm{L}} \in \mathbb{R}^{P \times d_\mathrm{m}}$. The detailed process is presented in Section \ref{sec:interface}.

\subsubsection{RGB Camera}
% To adapt the transformer architecture for computer vision tasks, we follow the standard approach of representing an image as a sequence of patches as the pre-processor. 
% An input RGB image of dimension ${H' \times W' \times 3}$ is first partitioned into a grid of $N_{\mathrm{C}}$ non-overlapping patches, each of size $r \times r$. 
% These patches are then flattened and linearly projected into $D$-dimensional embedding vectors. 
% This entire patch embedding process is efficiently implemented using a single 2D convolution with a kernel size and stride equal to the patch size $r$. 
% This procedure yields a sequence of $N_{\mathrm{C}}=(H' \cdot W')/r^2$ patch embeddings. 
% The dimension of each embedding vector is $C_{\mathrm{C}} = 3r^2$, resulting in an input sequence of dimension $N_{\mathrm{C}}\times C_{\mathrm{C}}$. 
The visual feature encoder is designed to extract high-dimensional semantic representations from raw camera observations. We employ a pre-trained vision transformer (ViT-B/16) \cite{dosovitskiy2020image} as the structural backbone.

To preserve the spatial integrity and aspect ratio of the input images, we implement a multi-stage preprocessing pipeline. Specifically, an input image is first isotropically resized such that its shorter dimension matches the model's input requirement, i.e., $224$ pixels. Subsequently, a center crop is applied to produce a fixed-size tensor $\mathbf{X}_{\mathrm{C}} \in \mathbb{R}^{3 \times 224 \times 224}$. This strategy minimizes geometric distortion compared to naive resizing, thereby maintaining the semantic consistency of local features.

The preprocessed image is partitioned into $N_{\mathrm{p}} = 224^2 / Z^2$ non-overlapping patches with a patch size of $Z$. Each patch is flattened and projected into a $d_{\mathrm{C}}=768$ dimensional latent space. Let $\boldsymbol{\Omega}_0$ be the input sequence to the transformer encoder, defined by
\begin{equation}
    \boldsymbol{\Omega}_0 = [\mathbf{x}_{\mathrm{cls}}; \mathbf{x}_1 \mathbf{E}_{\mathrm{p}}; \mathbf{x}_2 \mathbf{E}_{\mathrm{p}}; \dots; \mathbf{x}_{N_{\mathrm{p}}} \mathbf{E}_{\mathrm{p}}] + \mathbf{E}_{\mathrm{pos}},
\end{equation}
where $\mathbf{x}_{\mathrm{cls}}\in \mathbb{R}^{1 \times (3Z^2)}$ denotes the classification token, and $\mathbf{x}_i\in \mathbb{R}^{1 \times (3Z^2)}$ is the flattened patch. $\mathbf{E}_{\mathrm{p}} \in \mathbb{R}^{(3Z^2) \times d_{\mathrm{C}}}$ is the patch embedding projection and $\mathbf{E}_{\mathrm{pos}}\in \mathbb{R}^{(N_\mathrm{p}+1) \times d_{\mathrm{C}}}$ denotes the learnable positional embeddings.

The sequence, $\boldsymbol{\Omega}_0$, is then forwarded through the transformer encoder, which consists of $12$ layers of multi-head self-attention and multi-layer perceptron blocks. Let $\boldsymbol{\Omega} \in \mathbb{R}^{(N_{\mathrm{p}}+1) \times d_{\mathrm{C}}}$ denote the output hidden states from the final layer. 
Instead of utilizing the global class token, $\boldsymbol{\Omega}^{(0)}$, corresponding to $\mathbf{x}_{\mathrm{cls}}$ for image-level classification, we extract the sequence of transformed patch tokens to preserve spatial information. The resulting feature map, $\tilde{\mathbf{X}}_{\mathrm{C}} \in \mathbb{R}^{N_{\mathrm{p}} \times d_{\mathrm{C}}}$, is obtained by slicing the output tensor:
\begin{equation}
    \tilde{\mathbf{X}}_{\mathrm{C}} = \boldsymbol{\Omega}[2:N_{\mathrm{p}}+1, :].
\end{equation}

This feature map, $\tilde{\mathbf{X}}_{\mathrm{C}}$, provides a dense, spatially-aware representation of the scene, where each token encapsulates local semantic information conditioned on the global context through the self-attention mechanism.

For feature embedding, we adopt the same cross-attention mechanism as the LiDAR feature embedding for RGB images, along with the subsequent replication and concatenation process to transform the feature embeddings of the undersampled RGB images into historical image feature embedding $\mathbf{U}_{\mathrm{C}} \in \mathbb{R}^{P \times d_\mathrm{m}}$, as detailed in the following subsection.

\subsection{Modality Interface} \label{sec:interface}
% While the modality encoders extract tailored representations from heterogeneous data sources, LiDAR and camera features remain asynchronous due to the much lower data-sampling frequency compared with the beam index, and cannot be directly processed by an LLM that has been} pre-trained on natural language.
Although tailored representations are extracted from heterogeneous modalities, the inherent sampling rate disparity between the sensors and the beam index causes LiDAR and camera features to remain asynchronous with the beam index features. Furthermore, such non-textual representations cannot be directly ingested by a natural language-pretrained LLM.
To bridge this gap, the modality interface is designed to synchronize, fuse, and translate these disparate data streams. First, it addresses the challenge of different sensor sampling rates by temporally aligning the lower-frequency perceptual data to the high-frequency timeline of the communication channel. Subsequently, a cross-modality attention layer is employed to aggregate the complementary spatial and temporal messages from all aligned modalities. Finally, a modality reprogramming module translates the fused multimodal tensor into language-aligned embeddings. This vital transformation effectively projects the physical sensory data into a unified vocabulary space, enabling the pre-trained LLM to reason comprehensively about the multimodal environment.

\subsubsection{Temporal Alignment} 
A critical challenge in multimodal fusion is handling the disparate sampling rates of different sensors. Departing from conventional approaches that often downsample to the lowest common frequency---sacrificing temporal resolution---or simply ignore these asynchronies \cite{shibingpu, cuiyuanhao}, we introduce a strategy that preserves high-frequency information while respecting real-world data availability.

Our alignment methodology is anchored to the high-frequency timeline of the communication channel with a sampling period $T_{\mathrm{c}}$. Lower-frequency data from perceptual sensors, such as LiDAR and cameras with a period $T_{\mathrm{pc}}$, are up-sampled to match this reference timeline. Each sensor measurement corresponds to a window of $j = T_{\mathrm{pc}}/T_{\mathrm{c}}$ channel steps. Since a typical LiDAR sampling period of 100 ms or 50 ms is an integer multiple of the standard 10 ms 5G NR frame duration, this ratio naturally results in an integer value of $j=10$ or $j=5$, respectively. The cornerstone of our approach is the enforcement of causality: each sensor measurement is aligned to the end of its acquisition window, faithfully modeling real-world conditions and ensuring data is used after it becomes available.

\begin{figure}[t]
	\centerline{\includegraphics[width=3.4in]{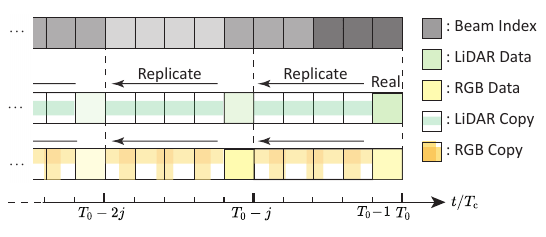}}
	\caption{Temporal alignment strategy for the modality interface.}
	\label{fig:align}
\end{figure}

This principle directly governs the construction of our multimodal input tensors. As illustrated in Fig.~\ref{fig:align}, for a history of $P$ time steps, the $\lfloor P/j \rfloor$ sparse encoded features for LiDAR and camera are temporally aligned with the high-frequency beam index sequence via backward replication. Specifically, each ``real'' feature available at every $j$-th time step is retrospectively copied to fill the $j-1$ preceding time steps within the same sampling interval. This process ensures that the resulting LiDAR sequence $\mathbf{U}_{\mathrm{L}} \in \mathbb{R}^{P \times d_{\mathrm{m}}}$ and camera sequence $\mathbf{U}_{\mathrm{C}} \in \mathbb{R}^{P \times d_{\mathrm{m}}}$ are temporally dense and synchronized with beam index sequence $\mathbf{U}_{\mathrm{B}}\in \mathbb{R}^{P \times d_{\mathrm{m}}}$ of length $P$. By expanding these feature embeddings along a newly introduced modality dimension, we obtain $\mathbf{U}'_{\mathrm{B}}$, $\mathbf{U}'_{\mathrm{L}}$, and $\mathbf{U}'_{\mathrm{C}} \in \mathbb{R}^{P \times 1 \times d_{\mathrm{m}}}$. Subsequently, concatenating these tensors along the modality dimension yields
\begin{equation}
    \mathbf{U}=\mathrm{Concat}(\mathbf{U}'_{\mathrm{B}},\mathbf{U}'_{\mathrm{L}},\mathbf{U}'_{\mathrm{C}}).
    \label{eq:final_concat}
\end{equation}
The obtained tensor, $\mathbf{U}\in\mathbb{R}^{P\times 3 \times d_{\mathrm{m}}}$, represents a temporally coherent and causally-sound fusion of all modalities, ready for processing by the subsequent cross-modality attention layers.

\subsubsection{Cross-Modality Attention} To analyze the features in $\mathbf{U}$ and aggregate messages from all modalities, we then employ a cross-modality attention layer. 
It utilizes a learnable query tensor $\mathbf{R}\in\mathbb{R}^{P\times 1 \times d_{\mathrm{m}}}$ to aggregate information, 
with $\mathbf{U}$ serving as both key and value:
\begin{equation}
    \mathbf{B}=\mathrm{CrossAttention}(\mathbf{R},\mathbf{U},\mathbf{U}).
\end{equation}
The resulting tensor, $\mathbf{B}\in\mathbb{R}^{P\times 1 \times d_{\mathrm{m}}}$, is a fused representation that is then passed to the next module. The layer itself is built upon a standard multi-head attention architecture, as discussed in Section~\ref{subsec:modalityencoder}.

\subsubsection{Modality Reprogramming} The fused feature tensor $\mathbf{B}$ is a powerful representation, which, however, cannot be directly processed by an LLM pre-trained on natural language. To bridge this gap, we introduce a modality reprogramming module that translates $\mathbf{B}$ into language-aligned embeddings, enabling the LLM to reason about the fused multimodal data.

This reprogramming is achieved by projecting $\mathbf{B}$ into a compact, continuous semantic subspace derived from the LLM's vocabulary. For computational efficiency, we avoid computing cross-attention over the entire vocabulary embedding matrix $\mathbf{E}\in \mathbb{R}^{V \times D}$, where $V$ and $D$ represent the vocabulary size and the hidden dimension of the LLM, respectively. Instead, we apply a learnable linear projection to $\mathbf{E}$ to construct a condensed set of latent text prototypes, $\mathbf{E}' \in \mathbb{R}^{V' \times D}$ (where $V' \ll V$). Unlike manually curated, discrete text prompts (e.g., "strong signal"), these data-driven virtual tokens act as learnable semantic anchors. This allows the model to automatically map complex, high-dimensional multimodal inputs to the most optimal language-aligned concepts during the end-to-end optimization process. Specifically, the translation of $\mathbf{B}$ is implemented with a multi-head cross-attention layer, where the text prototypes $\mathbf{E}'$ act as both key and value, and $\mathbf{B}'\in \mathbb{R}^{P \times d_\mathrm{m}}$ serves as the query, generated by squeezing the redundant modality dimension in $\mathbf{B}$. 
The operation is defined as
\begin{equation}
    \mathbf{Z} = \mathrm{CrossAttention}(\mathbf{B}', \mathbf{E}', \mathbf{E}'),
    \label{eq:reprogramming_attn}
\end{equation}
where the output $\mathbf{Z} \in \mathbb{R}^{P \times D}$ is the reprogrammed feature sequence. These vectors now reside in the same embedding space as the LLM's vocabulary, making them a suitable input for the LLM backbone.

% \textbf{Prompt-as-Prefix:} To guide the LLM's reasoning for the beam prediction task, we prepend a structured natural language prompt as a prefix to the reprogrammed feature sequence $\mathbf{Z}$. This prompt is a structured string that concatenates key information: a static dataset description, an explicit task instruction (e.g., ``forecast the next $W$ steps''), and instance-specific statistics such as minimum/maximum values and overall trend.

% This text is tokenized into prompt embeddings, $\mathbf{P}_{\text{prom}}$, which are then concatenated with the reprogrammed sequence $\mathbf{Z}$ to form the final input for the LLM as
% \begin{equation}
%     \tilde{\mathbf{Z}} = \mathrm{Concat}(\mathbf{P}_{\text{prom}}, \mathbf{Z}),
% \end{equation}
% which is subsequently processed by the LLM backbone.
\subsubsection{Prompt-as-Prefix}
To guide the LLM's reasoning process and anchor its general capabilities to the specifics of the beam prediction task, we construct a structured natural language prompt for each input sample according to \cite{bpllm}. 
This prompt is prepended as a prefix to the reprogrammed feature sequence, a technique we term prompt-as-prefix.
The combined sequence is then fed into the LLM backbone.

Each prompt is a structured string that concatenates several key pieces of information, designed to provide a comprehensive context for the model:
\begin{itemize}
    \item Dataset Description: A static text describing the overall dataset, designed to be: ``Beam prediction dataset is a collection of millisecond data from a multiple-input single-output system, describing the movement of beam direction over time, which typical remains constant within several time steps.''.
    \item Task Description: An explicit instruction defining the forecasting goal, such as ``forecast the next $W$ steps given the previous $P$ steps information,'' where $P$ is the look-back window and $W$ is the prediction horizon.
    \item Instance-Specific Statistics: A dynamic summary of the current input sequence's statistical properties, including: basic statistics like the minimum, maximum, and median values of the beam indices; a textual description of the overall {beam index} trend (either ``upward'' or ``downward''), determined by the sign of a trend metric; the top five most significant time intervals where the data shows the strongest repeating patterns, discovered by calculating how the time series correlates with its own past via fast Fourier transform.
    % the top five most significant temporal lags, identified by computing the autocorrelation function of the time series via fast Fourier transform.
\end{itemize}
% This structured text is then passed through the pre-trained LLM's tokenizer and embedding layer to generate a sequence of prompt embeddings, $\mathbf{P}_{\text{prom}} $ This ensures that the prompt resides in the same semantic space as the LLM's internal representations. Finally, the prompt embeddings are concatenated with the reprogrammed feature sequence $\mathbf{Z}$ along the temporal dimension to form :
% \begin{equation}
%     \tilde{\mathbf{Z}} = \text{Concat}(\mathbf{P}_{\text{prom}}, \mathbf{Z}),
%     \label{eq:concat}
% \end{equation}
% which is further processed by the LLM backbone.

This structured text is then passed through the pre-trained LLM's tokenizer and embedding layer to generate a sequence of prompt embeddings, $\mathbf{P}_{\text{prom}} \in \mathbb{R}^{L_{\text{prom}} \times D}$, where $L_{\text{prom}}$ denotes the token length of the prompt. 
This ensures that the prompt resides in the same semantic space as the LLM's internal representations. 
Finally, the prompt embeddings are concatenated with reprogrammed feature sequence $\mathbf{Z}$ to form unified input $\tilde{\mathbf{Z}} \in \mathbb{R}^{(L_{\text{prom}} + P) \times D}$:
\begin{equation}
    \tilde{\mathbf{Z}} = \mathrm{Concat}(\mathbf{P}_{\text{prom}}, \mathbf{Z}),
\end{equation}
which is further processed by the LLM backbone.

\subsection{Output Projection}
The output of the LLM backbone, denoted as $\tilde{\mathbf{O}}_{\text{LLM}} \in \mathbb{R}^{P' \times D}$, is a sequence of high-dimensional hidden states, where $P'$ is the length of the concatenated prompt and reprogrammed features. To transform these abstract representations into a concrete forecast, we design a dedicated prediction head.

First, we extract the feature vectors corresponding only to our original input sequence, discarding the outputs associated with the prompt prefix. 
This yields a matrix $\mathbf{O}_{\text{LLM}} \in \mathbb{R}^{P \times D}$. 
We then reduce the feature dimension from $D$ to a smaller, intermediate dimension $D_{\text{ff}}$ by selecting the initial components of each vector, which produces $\mathbf{O}'_{\text{LLM}}\in\mathbb{R}^{P\times D_{\mathrm{ff}}}$.

The core of the prediction head is a linear projection layer that directly maps the flattened sequence of features to the desired prediction horizon. 
Specifically, $\mathbf{O}'_{\text{LLM}}$ is flattened into a single long vector and then linearly projected to a vector of length $W$. This can be formulated as
\begin{equation}
    \hat{\mathbf{y}}_W = \text{Flatten}(\mathbf{O}'_{\text{LLM}})\cdot\mathbf{W}_{\text{out}} + \mathbf{b}_{\text{out}},
\end{equation}
where $\mathbf{W}_{\text{out}} \in \mathbb{R}^{(P \cdot D_\mathrm{ff}) \times W}$ and $\mathbf{b}_{\text{out}}\in \mathbb{R}^W$ are the learnable weights and bias of the final linear layer.

Finally, output $\hat{\mathbf{y}}_W$ is passed through a denormalization layer to revert the prediction to the original data scale, producing final forecast $\hat{\mathbf{q}}_W$. The loss function is designed to minimize the mean-squared error between predicted $\hat{\mathbf{q}}_W$ and ground truth $\mathbf{q}^{*}_W$, which can be expressed as

\begin{equation}
    \mathrm{Loss} = \frac{1}{W}\sum_{m=1}^W(\hat{q}_m-q^*_m)^2.
\end{equation}

\section{Multimodal-Wireless Dataset} \label{sec:dataset}
In this section, we briefly introduce Multimodal-Wireless, a large-scale open-source dataset for multimodal sensing and communication research. The dataset is collected in a dynamic vehicular environment, where vehicles and RSUs gather multimodal data for collaborative perception. Alongside conventional sensors (LiDAR, RGB and depth cameras, radar and IMU), both RSUs and vehicles use antenna arrays for communication. Specifically, the vehicles are modeled as users and the RSUs as BSs.

In what follows, we first give the general framework of the dataset to illustrate how we generate data. Then, we introduce the parametric details of all modalities in the dataset and how they correspond with our multimodal LLM-based beam prediction task. Notably, the following content serves as a brief introduction to the Multimodal-Wireless dataset, and readers are referred to \cite{icc2026} for a complete and detailed account. The dataset is publicly available in \url{https://le-liang.github.io/mmw/}.

\subsection{Generation Workflow}
The dataset’s generation pipeline combines CARLA \cite{carla}, the autonomous driving simulator, Sionna \cite{sionna}, the ray-tracing engine, and Blender \cite{blender}, the physical modeling software, to ensure spatial and temporal consistency across platforms. We utilize Blender to interface CARLA with Sionna’s ray-tracing engine, ensuring all modalities share a unified world. The process, as illustrated in Fig. \ref{fig:workflow}, comprises the following stages:
\begin{figure*}[t]
	\centerline{\includegraphics[width=6.6in]{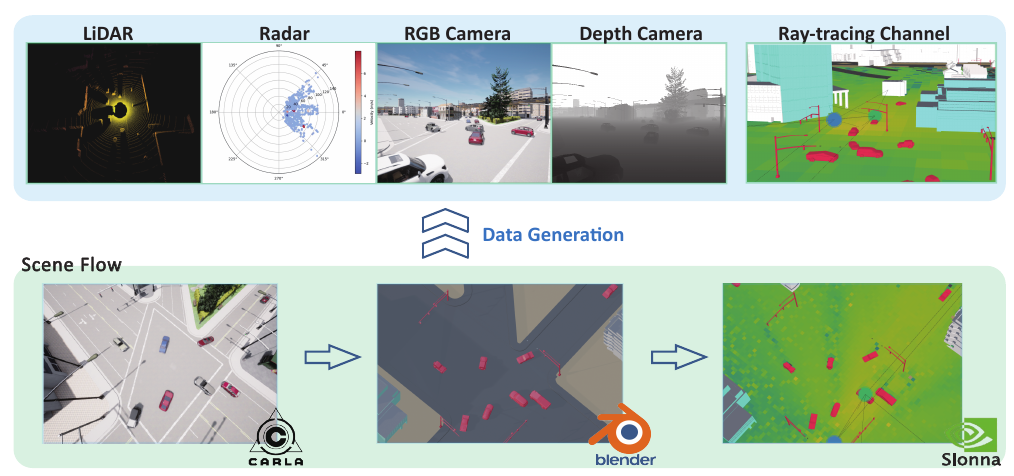}}
	\caption{Cross-platform data generation workflow for the Multimodal-Wireless dataset.}
	\label{fig:workflow}
\end{figure*}
\begin{itemize}
    \item \textbf{Scenario Execution and Data Capture in CARLA:} We first define and execute our scenario in CARLA. For each frame, five types of sensory data (LiDAR, RGB and depth camera, radar, and IMU) are captured. During the simulation, the dynamic state of each frame is recorded in a configuration file, consisting of the position and rotation of all vehicles in the scenario.
    \item \textbf{Scenario Reconstruction in Blender:} The static town map from CARLA is first established in Blender as a base environment. Then, for each frame, the pose information from its configuration file is used to place and orient all dynamic actors within this environment programmatically. Each fully constituted frame is then exported as a self-contained Sionna scene.
    \item \textbf{Channel Generation in Sionna:} Finally, the exported scenes are sequentially loaded into Sionna. The transmitter and receiver locations for each link are configured based on the pose data in the configuration files. Sionna's ray-tracing engine computes the detailed channel impulse response and path parameters, which constitute the communication modality for the Multimodal-Wireless dataset.
\end{itemize}

\subsection{Communication Modality}
With the dynamic scenarios fully reconstructed in a Sionna-compatible format, Sionna serves as the ray-tracing engine to generate the multi-path channel within the scenes from Blender. This subsection provides a detailed account of the technical procedures and the channel parameters involved.

In our primary setup, we consider the communication link from the RSU to the vehicles, which serve as the mobile users. Notably, this framework is inherently extensible to vehicle-to-vehicle communications. For the channel computation, we equip the RSU with a ULA of $N_{\mathrm{t}}$ transmit antennas and each vehicle with a ULA of $N_{\mathrm{r}}$ receive antennas for 2D settings and a uniform planar array (UPA) for 3D settings. 
After defining the antenna patterns and polarization, Sionna's ray-tracing engine is invoked to compute the propagation paths. 

To balance physical accuracy with computational feasibility, we collect the LOS path and all first-order reflection events. 
For each of the $L$ resulting paths, we store its fundamental physical properties: the azimuth and zenith AoD and angles-of-arrival AoA, ($\phi_{l}^{\mathrm{a}/\mathrm{z}}$, $\theta_{l}^{\mathrm{a}/\mathrm{z}}$), the propagation delay, $\tau_l$, and the complex gain matrix, $\mathbf{A}_l \in \mathbb{C}^{N_{\mathrm{r}}\times N_{\mathrm{t}}}$, representing the complex gain of the $l$-th path at the carrier frequency $f_{\mathrm{c}}$.

The fundamental multi-path parameters, especially ($\{\mathbf{A}_l\}_{l=1}^{L}, \{\tau_l\}_{l=1}^{L}$), allow us to generate custom single-carrier or multi-carrier (e.g., OFDM) frequency-domain channels by coherently summing the contributions of all paths at frequency $f_k$, according to the relationship:
\begin{equation}
\label{eq:Sionna_channel}
    \mathbf{H}(f_k) = \sum_{l=1}^{L} \mathbf{A}_{l} e^{-j2\pi f_k \tau_{l}},
\end{equation}
where $\mathbf{H}(f_k)$ is the channel frequency response at the $k$-th subcarrier.

All simulation parameters of Multimodal-Wireless are summarized in Table~\ref{tab:Sionna}. 
\begin{table}[t] 
    \centering
    \caption{Sionna Simulation Parameters in Multimodal-Wireless.}
    \label{tab:Sionna}
    \renewcommand{\arraystretch}{1.3}
    \begin{tabular}{|c|l|l|} % Changed first column to 'c' for better vertical centering
        \hline 
        \quad & \textbf{Parameter} & \textbf{Value} \\
        \hline
        % --- Fixed Parameters Section ---
        \multirow{5}{*}{\begin{turn}{90} \hspace{-1.5em} \textbf{Ray-tracing} \end{turn}} 
        & Carrier Frequency & $28$~GHz\;/\;$4.9$~GHz \\
        \cline{2-3}
        & Antenna Pattern & Dipole \\
        \cline{2-3}
        & Ray Samples Launched & $10^6$ \\
        \cline{2-3}
        & Maximum Reflection Order & $1$ \\
        \cline{2-3}
        & Polarization & Vertical \\
        \cline{2-3}
        & Frame Duration & $10$ ms  \\
        \hline
        % --- Flexible Parameters Section ---
        \multirow{6}{*}{\begin{turn}{90} \hspace{-1.5em} \textbf{Communication} \end{turn}} 
        & Subcarrier Spacing & $120$ kHz   \\
        \cline{2-3}
        & Number of Subcarriers & $1024$   \\
        \cline{2-3}
        & Transmit Array Size &  $1\!\times\!4/16/64/256$ (2D)\;/\;$8\!\times\!8$ (3D)\!\!\!\\
        \cline{2-3}
        & Receive Array Size & $1\!\times\!4/16$ (2D)\;/\;$8\!\times\!8$ (3D)   \\
        \cline{2-3}
        & Number of Transmit Antennas\!\!\! & $4/16/64/256$ (2D)\;/\;$64$ (3D)  \\
        \cline{2-3}
        & Number of Receive Antennas\!\!\! & $4/16$ (2D)\;/\;$64$ (3D)  \\
        \hline 
    \end{tabular}
\end{table}
We categorize them into two groups: \textbf{ray-tracing parameters}, which are set during the physics-based ray-tracing process, and \textbf{communication parameters}, which can be customized by the user when synthesizing the frequency-domain channel from the raw path data ($\{\mathbf{A}_l, \tau_l\}_{l=1}^{L}$). This mechanism grants users maximum control over the final channel realization.

Notably, we utilize 2D data collected with $1\times64$ ULA in the BS and $1\times16$ ULA in the user for model training and testing. Thus, this model aligns with the multi-path channel formulation in \eqref{eq:channel_model}, where the per-path matrix, $\mathbf{A}_{n,l}$, encapsulates the path gain and spatial signatures by
\begin{equation}
    \mathbf{A}_{l}=\sqrt{\frac{N_\mathrm{t}N_\mathrm{r}}{L}}\alpha_{l}\mathbf{a}_{\mathrm{r}}\left(\theta^{\mathrm{z}}_{l}\right)\mathbf{a}_{\mathrm{t}}^{\mathrm{H}}\left(\phi^{\mathrm{z}}_{l}\right),
\end{equation}
where $\theta^{\mathrm{z}}_{l}$ and $\phi^{\mathrm{z}}_{l}$ denote the dumped zenith AoA and AoD, respectively, corresponding to the 2D AoA and AoD of 2D scenarios in \eqref{eq:channel_model}.

\subsection{Sensor Modality}
In the CARLA simulator, all sensory data are collected at a base sampling rate of 100 Hz. This high-frequency data acquisition ensures that a detailed raw data stream is available, from which users are allowed to downsample to any practical frequency that is suitable for their specific application. In this subsection, we present the configuration of the sensors and the physics-based weather impairment modeling employed to enhance data fidelity.

\subsubsection{Sensor Configuration and Specifications}
Multimodal-Wireless utilizes two distinct sensor-equipped platforms: vehicles and RSUs. Each vehicle is equipped with a sensor suite that includes four RGB cameras. These cameras are positioned to face the front, back, left, and right, thereby providing full 360-degree visual coverage around the vehicle. Beyond the cameras, each vehicle is also equipped with a LiDAR sensor and an IMU. The RSUs are outfitted with a different set of sensors, with each unit containing an RGB camera, a depth camera, a LiDAR, and a radar. In our network, we utilize the LiDAR and RGB camera modality as sensing data.

\begin{figure}[t]
	\centerline{\includegraphics[width=3.4in]{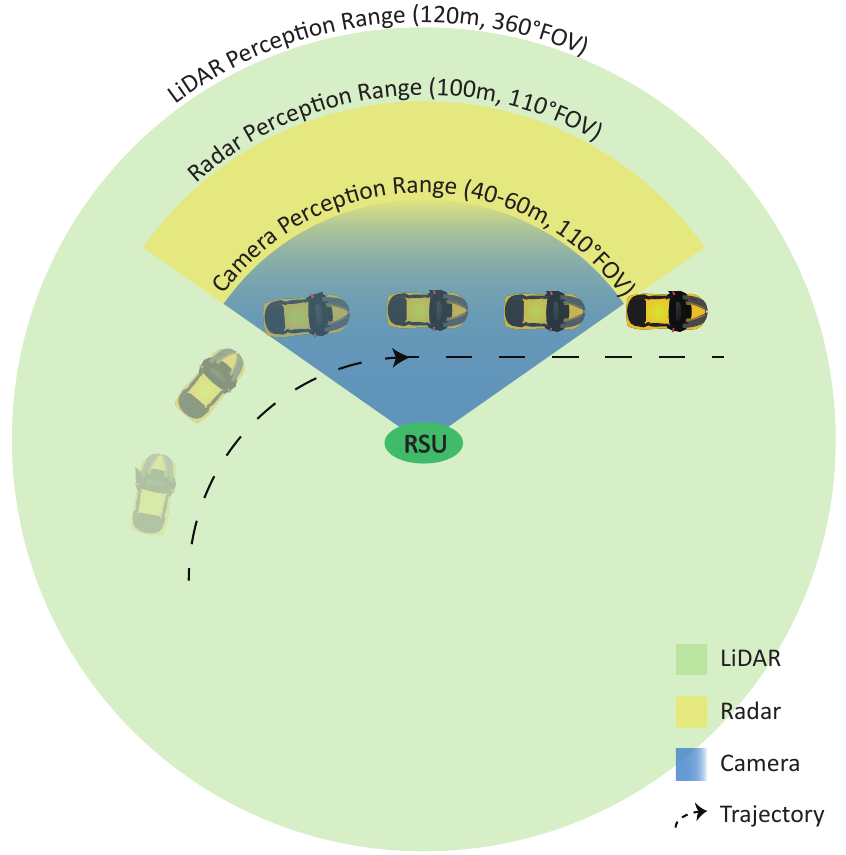}}
	\caption{Illustration of sensor perception ranges with vehicle trajectory.}
	\label{fig:range}
\end{figure}

However, the perception ranges of these sensor types vary significantly, as illustrated in Fig. \ref{fig:range}, which also shows a schematic trajectory of a vehicle moving through the environment. Typically, the effectiveness of the RSU camera and radar is constrained by a limited field of view (FOV). Consequently, these two sensors only detect the vehicle in a part of the overall scenario. In contrast, the LiDAR on the RSU can sense the vehicle across the entire area. The detailed technical specifications for all of these sensors are provided in Table \ref{tab:lidar}.

\subsubsection{LiDAR Weather Impairment Modeling}
While the CARLA simulator provides high-fidelity visual rendering for adverse weather, its native LiDAR sensor operates based on geometric ray-casting and does not physically interact with environmental particles. To bridge the gap between simulation and real-world sensing, we implement a physics-based post-processing pipeline to synthesize realistic weather-induced impairments on the raw LiDAR data.

\textbf{Fog Simulation:} Following the optical model in \cite{fog_ref}, we simulate the impact of fog through two mechanisms: \textit{hard target attenuation} and \textit{soft target backscattering}. Based on the Beer-Lambert law, the intensity of points reflected from solid objects is exponentially attenuated according to the distance and fog density. This attenuation process is formulated as
\begin{equation}
    I_{\mathrm{fog}} = I_{0} \cdot e^{-2\alpha_{\mathrm{fog}} R},
\end{equation}
where $I_{\mathrm{fog}}$ is the received intensity from the hard target in foggy conditions, $I_{0}$ represents the intensity in clear weather, $R$ denotes the radial distance to the target, and $\alpha_{\mathrm{fog}}$ is the fog extinction coefficient. The factor $2$ accounts for the round-trip path of the laser beam. Furthermore, to simulate the ``screen effect'' where fog itself reflects laser beams, we introduce phantom points (soft targets) by querying a pre-calculated backscattering intensity table. If the fog backscattering intensity, $I_{\mathrm{bck}}$, exceeds the attenuated object return $I_{\mathrm{fog}}$, the point is characterized as a fog point, introducing realistic spatial noise and occlusion.
% \textbf{Fog Simulation:} Following the optical model in \cite{fog_ref}, we simulate the impact of fog through two mechanisms: \textit{hard target attenuation} and \textit{soft target backscattering}. Based on the Beer-Lambert law, the intensity of points reflected from solid objects is exponentially attenuated according to the distance and fog density. Furthermore, to simulate the ``screen effect'' where fog itself reflects laser beams, we introduce phantom points (soft targets) by querying a pre-calculated backscattering intensity table. If the fog return intensity exceeds the attenuated object return, the point is characterized as a fog point, introducing realistic spatial noise and occlusion.

\textbf{Rain Simulation:} For rainy scenarios, we adopt the parametric model from \cite{rain_ref} to inject three distinct physical artifacts based on the rainfall rate $r$ (mm/h): \textit{distance perturbation}, \textit{intensity attenuation}, and \textit{point dropping}. To model the refraction and scattering variance, Gaussian noise is added to the radial distance of each point. The perturbed distance $R'$ is modeled as $R' = R + \epsilon$, where $\epsilon \sim \mathcal{N}(0, \sigma_0^2)$ represents the range error, with variance $\sigma_0^2$ scaling with $r$. Concurrently, signal intensity is attenuated to reflect energy loss through the rain medium. According to the empirical relationship between rain extinction coefficient $\alpha_{\mathrm{rain}}$ and rainfall rate $r$, the attenuation is formulated as:
\begin{equation}
    \alpha_{\mathrm{rain}} = \beta \cdot r^b, \quad I_{\mathrm{rain}} = I_{0} \cdot e^{-2\alpha_{\mathrm{rain}} R},
\end{equation}
where $\beta$ and $b$ are empirical constants depending on the laser wavelength, and $I_{\mathrm{rain}}$ denotes the reduced intensity in rain. Finally, a sensitivity threshold $I_{\mathrm{th}}$ is applied to filter out points with intensity falling below the detectable limit ($I_{\mathrm{rain}} < I_{\mathrm{th}}$), accurately reproducing the reduced detection range and sparsity observed in real-world rainy conditions.
% \textbf{Rain Simulation:} For rainy scenarios, we adopt the parametric model from \cite{rain_ref} to inject three distinct physical artifacts based on the rainfall rate: \textit{distance perturbation}, \textit{intensity attenuation}, and \textit{point dropping}. To model the refraction and scattering variance, Gaussian noise is added to the radial distance of each point, with the variance scaling with distance and rain intensity. Concurrently, signal intensity is attenuated to reflect energy loss through the rain medium. Finally, a sensitivity threshold is applied to filter out points with intensity falling below the detectable limit, accurately reproducing the reduced detection range and sparsity observed in real-world rainy conditions.

\begin{table}[h!] 
    \centering
    \caption{Sensor specifications.}
    \label{tab:lidar}
    \renewcommand{\arraystretch}{1.3}
    \begin{tabular}{|l|l|}
        \hline 
        \textbf{Sensors} & \textbf{Attributes} \\
        \hline
        RGB Camera & $640 \times 480$ resolution, $110^{\circ}$ FOV \\
        \hline
        Depth Camera & $640 \times 480$ resolution, $110^{\circ}$ FOV \\
        \hline
        LiDAR & \makecell[l]{64 channels, $30k$ points per sample, \\ 
                            $120$ m capturing range, $-25^{\circ}$ to $2^{\circ}$ \\ 
                            vertical FOV} \\
        \hline
        IMU & \makecell[l]{Gyroscope noise: mean $0.001$ rad/s, \\ 
                            standard deviation (std)
                            $0.002$ rad/s, \\ accelaration noise: std $0.1$ m/s$^2$} \\
        \hline
        Radar & \makecell[l]{$2k$ points per sample, $100$ m capturing  \\ 
                            range, $30^{\circ}$ vertical FOV and $110^{\circ}$ 
                            \\ horizontal  FOV} \\
        \hline 
    \end{tabular}
\end{table}

To conclude, Multimodal-Wireless provides a high-fidelity, strictly synchronized multimodal testing environment by integrating CARLA, Blender, and Sionna. By offering accurate ray-tracing channels paired with weather-impaired sensory data, it overcomes the limitations of existing datasets. This highly extensible and realistic dataset provides the essential groundwork for training our multimodal LLM and validating its robustness under diverse dynamic scenarios.

\section{Simulation Results} \label{sec:simulation}
\subsection{Dataset and Setup}
We conduct our experiments using the Multimodal-Wireless dataset. It provides synchronized data streams from multiple sensors, including LiDAR and camera, along with communication channel information generated via precise ray-tracing. For our study, we primarily utilize the data collected under the ``sunny'' weather scenario. 
We utilize a total of 50 trajectories, 40 used for training ($41,800$ samples), 5 for validation ($6,000$ samples), and the remaining 5 for testing ($6,000$ samples). By strictly separating the trajectories across these sets, we ensure a disjoint split, meaning all reported results reflect the model's generalization capability in out-of-distribution settings.
The BSs are equipped with $1\times64$ ULAs, setting $N_{\mathrm{t}}$ to be $64$, whereas the users are equipped with $1\times16$ ULAs, with $N_{\mathrm{r}}=16$. The size of the beam codebook $Q$ for the BSs and users is set to 64. 
The time resolution for each prediction step is $10$ ms. 
For user mobility, the vehicle speed is not constant but varies adaptively according to traffic conditions, ranging from static positioning in traffic jams to high-speed cruising on clear avenues. This variability helps mitigate the risk of overfitting and endows the model with inherent robustness to speed fluctuations.

The system environment of the server is Ubuntu 22.04, with Python 3.8. The graphics card utilized is the NVIDIA GeForce RTX 4090 24 GB GPU.

\subsection{Model and Baselines}
We configure our model with a look-back window of $P=40$ time steps (i.e., $400$ ms) and a prediction horizon of $W=10$ time steps (i.e., $100$ ms). For the core of our architecture, we employ a pre-trained GPT-2 model \cite{gpt} as the LLM backbone. This choice strikes a balance between robust reasoning capabilities and practical inference speed. To demonstrate the practical superiority of our approach, we compare our proposed multimodal model—which relies on readily accessible, on-site LiDAR and camera data—directly against the original baseline from [5], which assumes the availability of perfect AoD knowledge. 
%We compare our multimodal approach against the state-of-the-art LLM-based method from \cite{bpllm}. The original baseline model utilizes historical beam indices and AoD as its inputs. To ensure a fair and practical comparison, we adapt this baseline by replacing the BS-unavailable AoD input with the readily accessible, on-site LiDAR and camera data processed by our proposed feature embedding modules.

\subsection{Numerical Results}
\begin{figure}[t]
	\centerline{\includegraphics[width=3.4in]{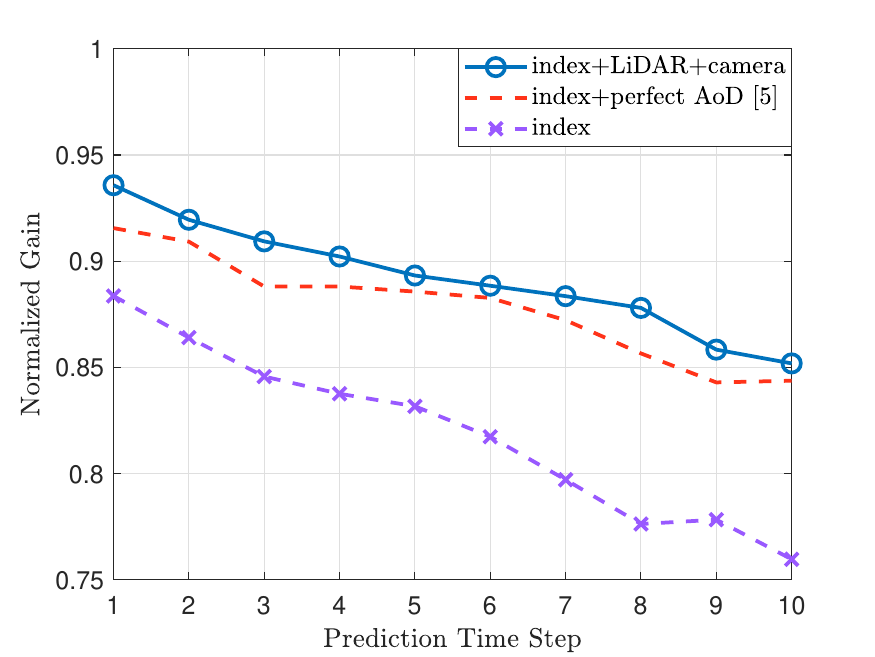}}
	\caption{Prediction performance of the proposed multimodal LLM-based method compared with the method in \cite{bpllm}.}
	\label{fig:main}
\end{figure}
Fig.~\ref{fig:main} compares the performance of the proposed multimodal LLM-based method with the LLM-based approach in \cite{bpllm}, evaluated in terms of the normalized gain defined in \eqref{eq:normalizedgain}. The baseline model is trained on the same dataset, leveraging beam indices and perfect AoD for long-term beam prediction. A practical baseline trained with only historical indices is also included. As shown in Fig.~\ref{fig:main}, by incorporating LiDAR and camera data, our multimodal LLM-based method consistently achieves close normalized gain with the AoD-based method in \cite{bpllm} across all prediction time steps, effectively solving the severe performance degradation caused by the absence of practically unavailable perfect AoD. 
% Specifically, the signal power after beamforming increases by an average of 0.15 dB for the final five time steps. 
This improvement arises because historical LiDAR and camera data provide valuable environmental context, allowing the model to better capture scene dynamics such as incoming traffic jams and more accurately predict subsequent indices.

% Then we challenged the model with a modified test set with NLOS conditions. While the original dataset is composed entirely of LOS channels, we synthetically created NLOS scenarios by removing the LOS path from 21\% of the test channels. As shown in Fig.~\ref{fig:nlos}, introducing NLOS paths distinguishes the two methods. The performance of the 'index+perfect AoD' method deteriorates sharply for longer prediction horizons. Conversely, the `index+LiDAR' method maintains high accuracy, demonstrating robustness. The performance gap widens as the prediction timestep grows,
% with the post-beamforming signal power increasing by an average of 0.18 dB for the final five time steps, 
% reflecting the value of LiDAR's contextual geometric information for capturing NLOS effects that cannot be extrapolated from AoD alone.
% \begin{figure}[t]
% 	\centerline{\includegraphics[width=3.4in]{fig/nlos.pdf}}
% 	\caption{Prediction performance in the test set with NLOS channels.}
% 	\label{fig:nlos}
% \end{figure}

\begin{figure}[t]
	\centerline{\includegraphics[width=3.4in]{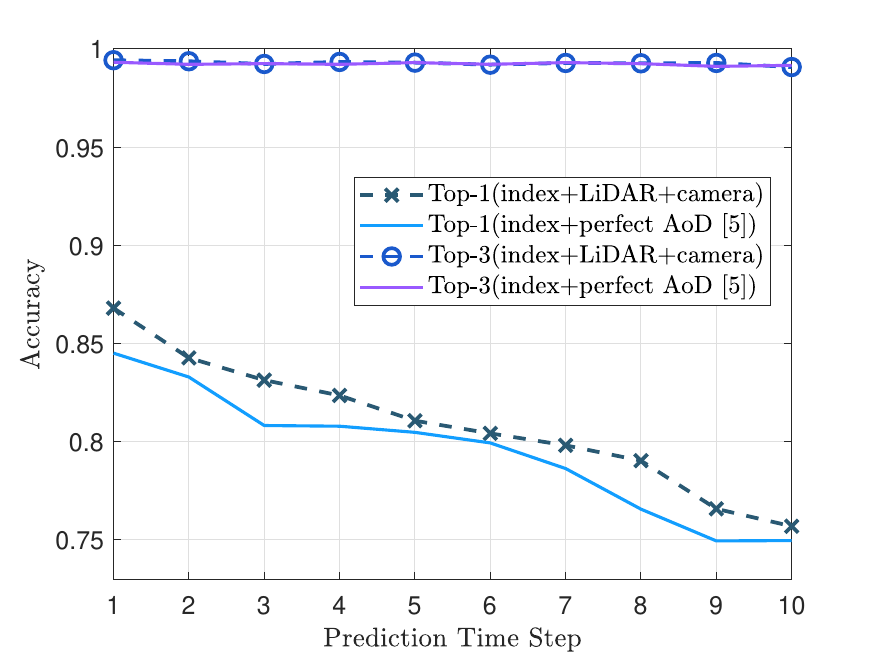}}
	\caption{Top-1 and Top-3 prediction accuracy of the proposed multimodal LLM-based method compared with the method in \cite{bpllm}}.
	\label{fig:topk}
\end{figure}

To evaluate the model performance from a classification perspective, we employ the Top-$k$ beam accuracy, which measures the probability that the predicted beam index, $\hat{q}_m$, is successfully included within the set of $k$ candidates with the highest normalized gains, denoted as $\mathcal{S}^{(k)}_{m,i}$. Mathematically, it is defined as:
\begin{equation}
    \text{Acc}_m@k = \frac{1}{N_{\text{test}}} \sum_{i=1}^{N_{\text{test}}} \mathds{1}(\hat{q}_m \in \mathcal{S}^{(k)}_{m,i}),
\end{equation}
where $\mathds{1}(\cdot)$ is the indicator function, which equals 1 if the condition is true and 0 otherwise.
As illustrated in Fig. \ref{fig:topk}, we compare the Top-1 and Top-3 accuracy of our multimodal method against the baseline with perfect historical AoD. First, the Top-3 accuracy for both methods remains consistently near $1.0$ across the entire prediction horizon. This indicates that even when the model misses the exact optimal beam (causing the decay in the Top-1 curve), the true optimal beam or a highly correlated neighbor is almost always present in the Top-3 suggestions. This suggests that the uncertainty of the LLM is well-confined within a narrow angular sector.
Second, our multimodal method consistently outperforms the AoD-based baseline in the Top-1 regime, particularly at later time steps (e.g., $m=10$). This reinforces the conclusion that multimodal sensory data provides superior context for anticipating complex user mobility compared to historical channel geometry alone.
In practice, this implies that by slightly increasing the overhead to sweep just 3 predicted beams, the system can eliminate the performance degradation caused by prediction errors, ensuring highly reliable connectivity.

We benchmark our approach against LSTM \cite{lstm} and ResNet \cite{resnet} baselines, both scaled up to match the parameter magnitude of the proposed LLM to ensure fairness. Table \ref{tab:model_comparison} reveals a compelling finding: despite our GPT2-based model incurring the highest computational complexity in terms of floating point operations (FLOPs), it demonstrates superior inference speed compared to the baselines. This efficiency stems from the highly parallel nature of the Transformer architecture, which effectively utilizes GPU resources. For real-time deployment, although the inference latency exceeds the single-step interval of 10 ms, this can be addressed via pipeline optimization or by directly leveraging the prediction horizon (e.g., applying predictions for the latter 5 steps). In terms of prediction quality, our multimodal model significantly outperforms others in Top-1 accuracy and normalized gain. This performance advantage is particularly evident in handling the dynamic vehicle speeds in our dataset, confirming the superior robustness of the LLM framework over traditional architectures.

\begin{table*}[t]
    \centering
    % 保持行高设置
    \renewcommand{\arraystretch}{1.4}
    
    % Caption中解释缩写
    \caption{Model complexity Analysis and Performance Comparison. \\Abbreviations: \textbf{Param.} (Parameters),  \textbf{Acc.} (Accuracy), \textbf{Avg.} (Average) and \textbf{Norm.} (Normalized).}
    
    % 【修改处】将第2列的 'c' 改为 'l'，使 Index/Cite/Ours 左对齐
    \begin{tabular}{|l|l|c|c|c|c|c|}
        \hline
        % 表头使用缩写
        \multicolumn{2}{|l|}{\textbf{Model}} & \textbf{Param.~(M)} & \textbf{FLOPs~(G)} & \textbf{Inference Time~(ms)} & \textbf{Avg. Top-1 Acc.} & \textbf{Avg. Norm. Gain}  \\
        \hline
        
        \multicolumn{2}{|l|}{LSTM} & 565.7 & 379.5 & 99.7 & 42.0\% & 51.3\%  \\
        \hline
        
        \multicolumn{2}{|l|}{ResNet} & 460.6 & 309.5 & 46.1 & 76.9\% & 85.3\%  \\
        \hline
        
        % GPT2 在第1列(l)，Index/Cite/Ours 在第2列(现已改为l)
        \multirow{3}{*}{GPT2} & Index & 754.1 & 488.1 & 34.2 & 72.8\% & 81.9\% \\
        \cline{2-7}
         & \cite{bpllm} & 755.9 & 488.1 & 33.1 & 79.5\% & 87.8\%  \\
        \cline{2-7}
         & Ours & 777.0 & 518.6 & 43.5 & \textbf{80.8\%} & \textbf{89.1\%}  \\
        \hline
    \end{tabular}
    \label{tab:model_comparison}
\end{table*}

\begin{figure}[t]
	\centerline{\includegraphics[width=3.4in]{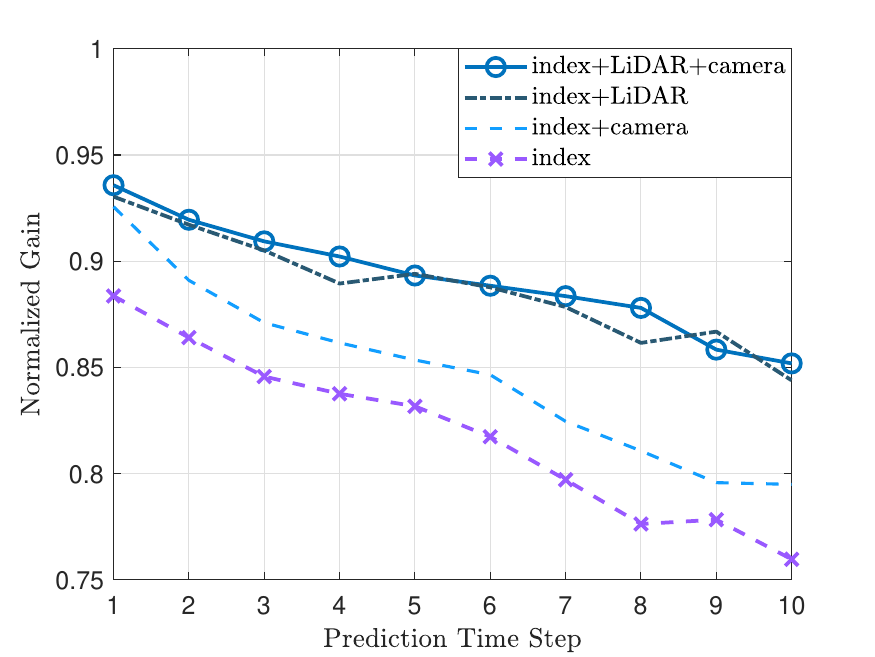}}
	\caption{Ablation study on different input modalities based on normalized gain.}
	\label{fig:ablation}
\end{figure}

For ablation study, we first investigate the contribution of different input modalities. As illustrated in Fig.~\ref{fig:ablation}, combining LiDAR with beam index data achieves performance close to the full-modality model, corresponding to the ``index+LiDAR'' curve. In contrast, removing LiDAR while retaining the camera input leads to a substantial degradation, reflected in the ``index+camera'' curve. This is due to the limited FOV of the RGB camera, which compromises the quality of the camera modality data. Notably, compared with the index-only baseline, the addition of the camera modality still improves performance, highlighting the effectiveness of our camera modality processing scheme.

\begin{figure*}[t]
	\centerline{\includegraphics[width=7.0in]{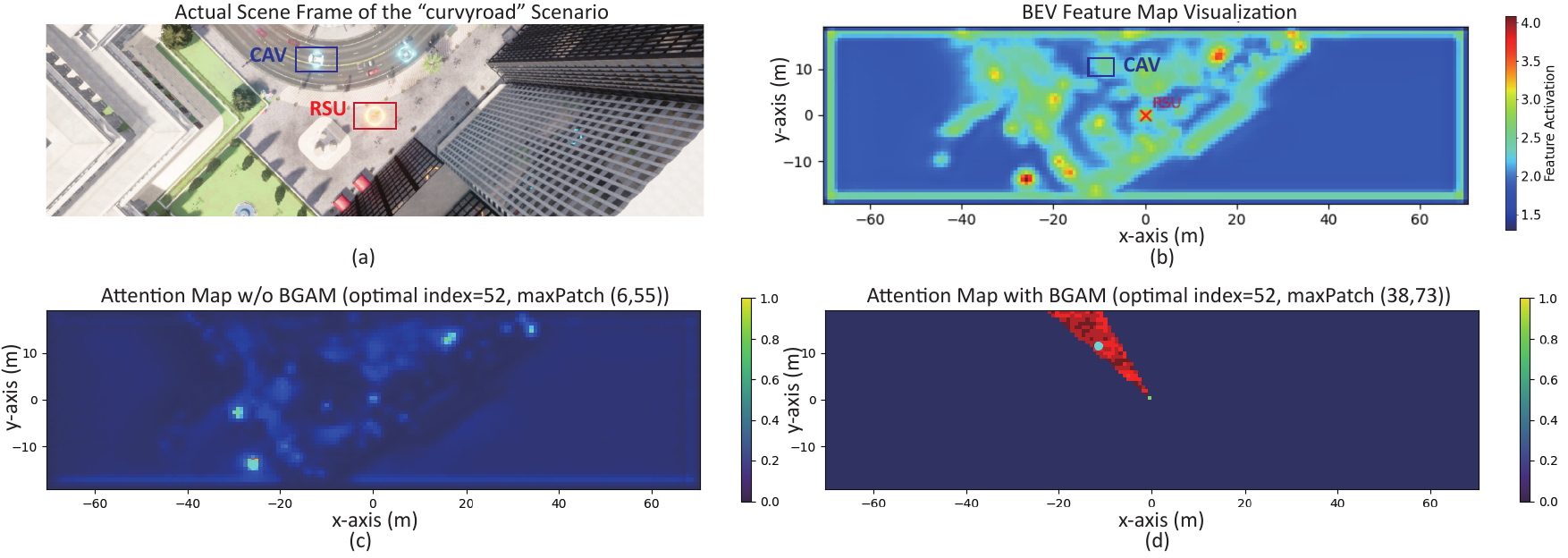}}
	\caption{Visualization of scene geometry and the effectiveness of BGAM in spatial disambiguation: (a) actual scene frame of the "curvyroad" scenario with User and RSU annotations; (b) corresponding LiDAR-based BEV feature map; (c) attention map without BGAM, displaying diffuse focus and background interference; (d) attention map with BGAM, showing precise, target-specific focus guided by historical beam indices.}
	\label{fig:bgam}
\end{figure*}

\begin{figure}[t]
	\centerline{\includegraphics[width=3.4in]{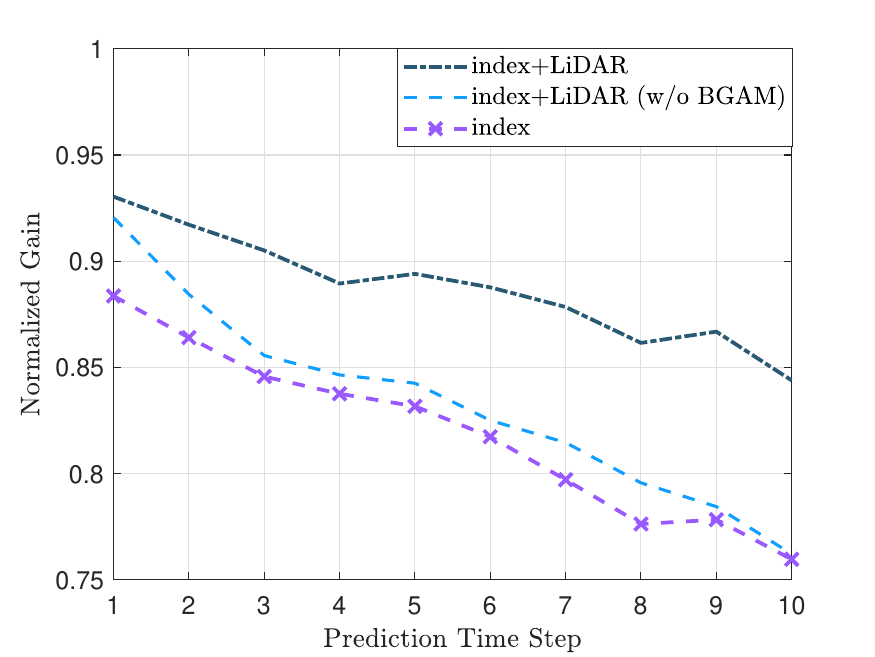}}
	\caption{Ablation study on BGAM based on normalized gain.}
	\label{fig:abla_bgam}
\end{figure}

We then perform an ablation study to investigate the effectiveness of the proposed BGAM method. Fig.~\ref{fig:bgam} visualizes the impact of BGAM on the model’s spatial attention performed in \eqref{eq:cross_attention} within the "curvyroad" scenario. As shown in Fig.~\ref{fig:bgam}(c), without BGAM, the attention map is highly diffuse, with the model focusing on prominent but irrelevant background structures. This lack of spatial guidance leads to a significant misalignment between the maximum attention patch and the actual location of the user, highlighting the difficulty in disambiguating the target from environmental noise. In contrast, the integration of BGAM, as illustrated in Fig.~\ref{fig:bgam}(d), introduces a directional prior derived from historical beam indices, effectively constraining the attention within a focused angular sector. This mechanism successfully suppresses interference from irrelevant spatial features and accurately localizes the user, as evidenced by the alignment of the peak attention patch with the user’s position in the BEV feature map. Moreover, as shown in Fig.~\ref{fig:abla_bgam}, removing BGAM leads to a substantial drop in normalized gain, yielding results comparable to the index-only model. This observation underscores the critical role of the BGAM method.

\begin{figure*}[t]
	\centerline{\includegraphics[width=6.6in]{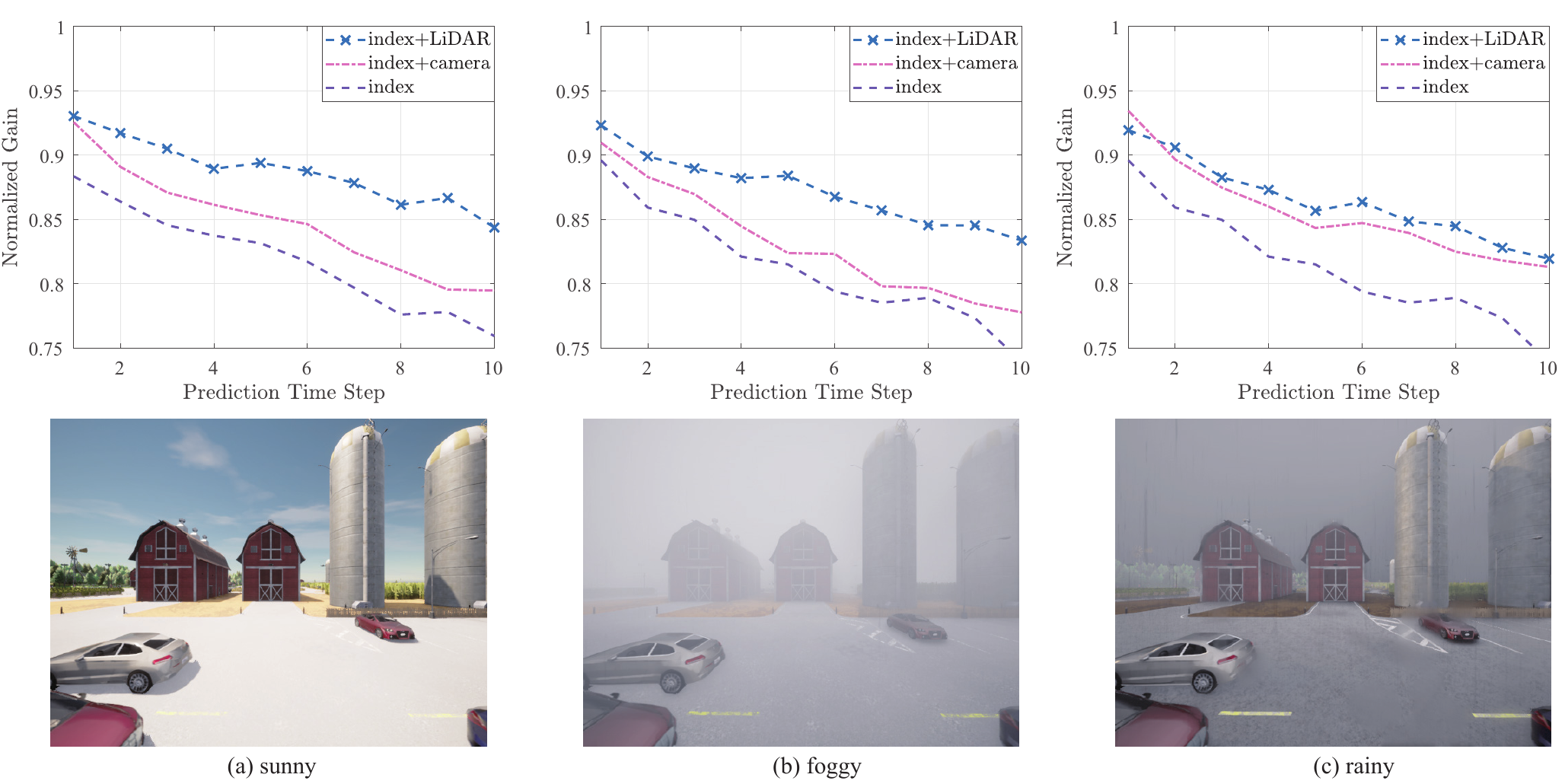}}
	\caption{Prediction performance and scene visualization under different weather conditions: (a) sunny, (b) foggy, and (c) rainy.}
	\label{fig:weathers}
\end{figure*}

Fig.~\ref{fig:weathers} shows the normalized gain evaluated on datasets with different weather scenarios, including sunny, heavy fog, and heavy rain conditions. Under sunny conditions, the camera modality provides clear visual information and leads to noticeable performance improvements. The LiDAR modality, unaffected by false echoes, delivers accurate environmental perception and significantly enhances the normalized gain. In contrast, under heavy fog, camera images become blurred and ambiguous, resulting in a small contribution. Moreover, water vapor in the air induces Mie scattering \cite{mie} and partial occlusion of laser beams, which attenuates return intensity and range while introducing short-range false echoes, thereby producing sparser and noisier LiDAR point clouds and reducing its effectiveness. However, LiDAR still provides a measurable performance gain. Under heavy rain, the images appear clearer yet darker due to the cloudiness, resulting in a further performance improvement. Meanwhile, the dense raindrops intensify false LiDAR echoes, further diminishing the performance benefit brought by LiDAR.

\begin{figure}[t]
	\centerline{\includegraphics[width=3.4in]{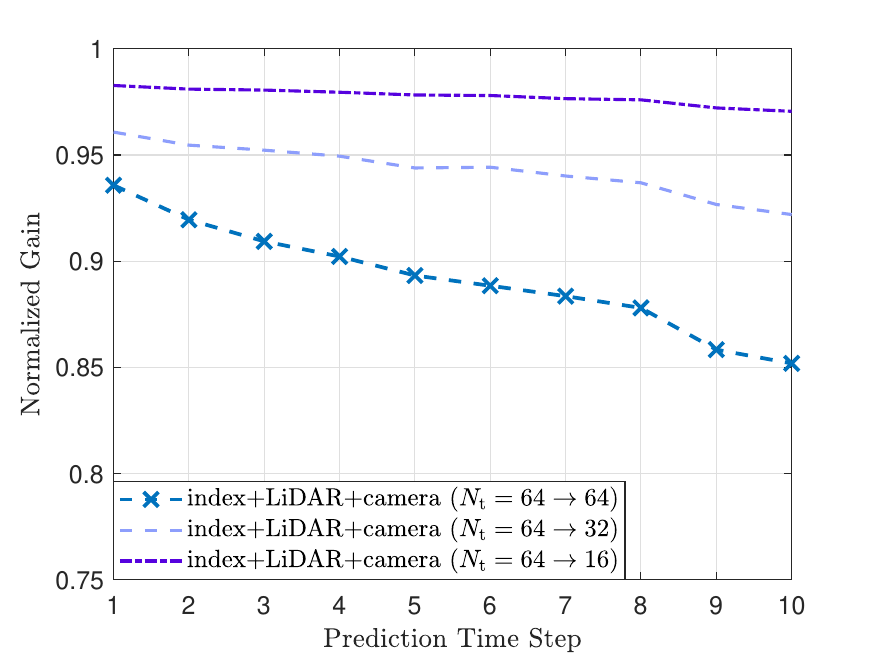}}
	\caption{Model robustness to mismatch of transmit antenna number $N_{\mathrm{t}}$.}
	\label{fig:robust_nt}
\end{figure}

Finally, we evaluate the robustness of our proposed model trained with $N_{\mathrm{t}}=64$ and codebook size $Q=64$ by testing the model against datasets with mismatched $N_{\mathrm{t}}$ configurations, revealing the model's excellent generalization capabilities. We challenge the model with test sets generated with $N_{\mathrm{t}}=16$ and $N_{\mathrm{t}}=32$. The results, presented in Fig.~\ref{fig:robust_nt}, confirm the model's robustness to antenna configuration mismatches given that the model trained in $N_{\mathrm{t}}=64$ adapts effectively to test sets with $N_{\mathrm{t}}=16$ and $32$. Moreover, the observed performance improvement with smaller $N_{\mathrm{t}}$ is expected since wider beams reduce the gain disparity between optimal and suboptimal beams.
% A larger $Q$ provides a finer quantization of the angular domain, making it easier to select a beam closely aligned with the optimal direction. This explains why the performance improves when testing on a larger codebook (e.g., the `64$\rightarrow$128' case), as it achieves a higher normalized gain.

\section{Conclusion} \label{sec:conclusion}
In this study, we have proposed a novel multimodal LLM-based framework for context-aware beam prediction that relies exclusively on readily available BS-side sensory data (LiDAR and RGB camera). The framework introduces two key technical innovations: (1) the BGAM mechanism, which leverages historical beam indices to dynamically guide the LiDAR encoder to focus on spatially relevant geometric features; and (2) a high-frequency temporal alignment strategy that upsamples sensory features to preserve channel temporal resolution (100 Hz) while ensuring causality. Furthermore, we construct Multimodal-Wireless, a large-scale, high-fidelity dataset generating synchronized channel and sensing data via ray-tracing across diverse weather conditions (sunny, rainy, foggy). Simulation results demonstrate that our approach significantly outperforms state-of-the-art LLM-based baselines that assume perfect AoD knowledge, exhibiting superior robustness against adverse weather and hardware configuration mismatches.

\vfill

\end{document}